\documentclass[review,12pt,authoryear]{elsarticle}

\usepackage{amssymb}
\usepackage{amsmath}
\usepackage{booktabs}
\usepackage{multirow}
\usepackage{tabularx}
\usepackage{array}
\usepackage{url}
\usepackage{soul}
\usepackage{tikz}
\usepackage[section]{placeins}
\usetikzlibrary{arrows.meta, positioning, shapes.geometric, calc}

\journal{Computers and Electronics in Agriculture}

\begin{document}

\begin{frontmatter}

%% Title
\title{
Non-Destructive Prediction of Fruit Ripeness and Firmness Using Hyperspectral Imaging and Lightweight Machine Learning Models
%Non-Destructive Fruit Ripeness and Firmness Prediction from Hyperspectral Imaging: Benchmarking Preprocessing Strategies and Machine Learning Algorithms %A Lightweight and Effective Machine Learning Approach 
}

%% Authors
\author[1]{Phongsakon Mark Konrad}
%\ead{phkon23@student.sdu.dk}

\author[2]{Casper Kunstmann-Olsen}

\author[2]{Jacek Fiutowski}

\author[1]{Serkan Ayvaz\corref{cor1}}
\ead{seay@mmmi.sdu.dk}
\cortext[cor1]{Corresponding author}

%% Author affiliations
\affiliation[1]{organization={The Maersk Mc-Kinney Moller Institute},%Department and Organization
            addressline={SDU Centre for Industrial Software, University of Southern Denmark},
            city={Sønderborg},
            postcode={6400},
            country={Denmark}}

\affiliation[2]{organization={Mads Clausen Institute},%Department and Organization
            addressline={SDU NanoSyd, University of Southern Denmark},
            city={Sønderborg},
            postcode={6400},
            country={Denmark}}

%\affiliation[3]{organization={Mads Clausen Institute},%Department and Organization
%            addressline={SDU Climate Cluster, University of Southern Denmark},
 %           city={Sønderborg},
 %           postcode={6400},
 %           country={Denmark}}

%% Abstract
\begin{abstract}
Post-harvest fruit quality assessment is essential for reducing food waste, yet reliable non-destructive methods typically depend on expensive hyperspectral cameras and computationally intensive deep learning models. These systems typically require GPU resources, large-scale training data, and domain expertise, limiting their feasibility for many real-world agricultural settings. This study systematically evaluates 20 classical machine learning algorithms on hyperspectral imaging data for simultaneous ripeness classification and firmness prediction across five fruit species, using cross-validated experimental design with Bayesian hyperparameter optimization. Data preprocessing strategy, particularly class balancing and spectral transformations, contributes as much to prediction accuracy as algorithm choice. Our results show that tree-based machine learning models can outperform state-of-the-art deep learning models reported in Fruit-HSNet. Moreover, the findings indicate that only three visible-range wavelengths are needed to recover over 94\% of full-spectrum accuracy, demonstrating that low-cost multispectral sensors combined with lightweight machine learning models can serve as practical alternatives to expensive hyperspectral cameras and complex deep learning approaches for practical fruit quality sorting.
\end{abstract}

%%Research highlights
%\begin{highlights}
%\item Only three visible-range wavelengths are needed for effective fruit quality sorting, enabling low-cost multispectral sensor deployment
%\item First systematic benchmark of lightweight ML  with preprocessing ablations for fruit quality prediction
%\item Non-destructive simultaneous ripeness and firmness prediction demonstrated across five fruit species on consumer CPU hardware
%\item Firmness prediction is inherently more accurate than ripeness classification, informing sensor and label design for post-harvest systems
%\item Data preprocessing choices rival algorithm selection in determining fruit quality prediction accuracy

%\end{highlights}

%% Keywords

\end{frontmatter}

\section{Introduction}
\label{sec:intro}

Commercial fruit quality assessment relies on destructive sampling of small fractions of harvests or subjective visual inspection \citep{lu2020recent}. A typical processing facility handles tens of thousands of items daily but can destructively test perhaps 100 samples. The remainder is graded by surface appearance, a weak proxy for internal quality that leads to misclassification, post-harvest losses, and inefficient supply-chain decisions.

Hyperspectral imaging (HSI) addresses this gap through non-destructive, contact-free measurement of wavelength-dependent reflectance across the visible and near-infrared spectrum. Spectral signatures encode biochemical and structural changes associated with ripeness, firmness, and spoilage \citep{yang2025hyperspectral}. Laboratory studies report accuracy exceeding 84\% for strawberry maturity assessment \citep{su2021strawberry}, yet commercial adoption remains limited.

Deep learning architectures achieve strong laboratory performance: 3D-CNNs reach 84\% accuracy for strawberry classification \citep{su2021strawberry}, transformer models reach 99.4\% for contamination detection \citep{guo2024daasst}, and real-time systems achieve 98.6\% accuracy \citep{gao2020realtime}. However, these approaches require GPU acceleration, large training datasets, and specialist expertise that most agricultural operations cannot justify. Classical machine learning offers a different set of tradeoffs. These methods train efficiently on limited data, run on consumer-grade CPU hardware, and produce interpretable feature importance scores amenable to validation against plant physiology. Whether classical methods can achieve competitive accuracy on hyperspectral fruit quality benchmarks remains uncharacterized.

Systematic benchmarking is lacking. Published comparisons typically fix a single preprocessing pipeline without ablation, leaving open the question of how much accuracy variation stems from data engineering rather than algorithm selection.

This study benchmarks 20 classical and gradient-boosted machine learning algorithms on paired ripeness and firmness prediction using the DeepHS Fruit dataset \citep{varga2021measuring}. A six-phase experimental design spans 200 model--configuration pairs, 4{,}000 Bayesian optimization trials, 10-fold cross-validation, ensemble learning, and explainable AI analysis. The central question is whether preprocessing decisions contribute as much to classification accuracy as algorithm choice, and if so, what mechanisms drive that effect. By evaluating all 20 algorithms under identical, controlled conditions, this benchmark provides a transparent reference point for the field. The main contributions are:

\begin{enumerate}
\item Systematic ablation across 200 configurations provides the first evidence that preprocessing shifts accuracy as much as algorithm selection (25.2 vs.\ 32.4~pp; bootstrap 95\% CI on the difference contains zero).
\item Firmness prediction outperforms ripeness classification across all 20 models ($d{=}1.84$), traced to label-structure asymmetry rather than algorithm-specific effects.
\item PCA degrades performance after engineered spectral transforms because variance-maximizing compression discards low-variance features that carry discriminative information.
\item Consensus feature ranking identifies three visible-range wavelengths (448, 540, 640\,nm) that recover 94.7\% of full-spectrum accuracy; approximate RGB wavelengths (450, 550, 651\,nm) perform comparably (96.1\%), indicating that band count rather than precise band placement drives the reduced-band performance.
\end{enumerate}

The remainder of this paper is organized as follows. Section~\ref{sec:related} reviews prior work on HSI-based fruit quality assessment, preprocessing pipelines, and explainability methods. Section~\ref{sec:methods} describes the dataset, feature engineering, and six-phase experimental design. Section~\ref{sec:results} presents results across all phases, including ablation, cross-validation, and XAI analysis. Section~\ref{sec:discussion} interprets the key findings and discusses limitations. Section~\ref{sec:conclusion} summarizes the contributions and outlines directions for future work.

\section{Related Work}
\label{sec:related}

Three areas of prior work frame the contribution of this study.

\subsection{HSI and Machine Learning for Fruit Quality}

Hyperspectral systems acquire spectral information across hundreds of narrow wavelength bands, spanning visible (400--700\,nm) and near-infrared (700--2500\,nm) regions. This spectral data enables detection of biochemical changes during fruit maturation: chlorophyll degradation, carotenoid accumulation, and cellular structure modifications \citep{lu2020recent}. Spectral-biochemical relationships have been established for quality attributes: chlorophyll absorption peaks around 680\,nm indicate ripeness state, while near-infrared signatures (850--950\,nm) correlate with cellular structure and firmness \citep{khodabakhshian2017application, feng2023nondestructive}. Translating these laboratory findings into deployable systems remains difficult.

Traditional machine learning approaches, including Partial Least Squares Regression (PLSR), Support Vector Machines (SVM), and tree-based ensembles, achieved 70--85\% accuracy with advantages in interpretability and computational efficiency \citep{lu2017innovative, khodabakhshian2017application}. Reviews of the field identify PLSR and SVM as the most commonly applied methods \citep{wieme2022biosystems}, and targeted wavelength selection can maintain classification accuracy with as few as six bands \citep{nagasubramanian2018genetic}. These practical insights have received limited attention as the field shifted toward deep learning, and systematic evaluation of traditional methods under standardized conditions remains sparse.

Despite these accuracy gains, deep learning models require GPU acceleration, specialized expertise, and extensive training datasets unavailable for diverse fruit varieties or regional conditions. Computational requirements exceed what most agricultural facilities can justify. Hybrid approaches combining traditional feature engineering with deep learning \citep{liu2024transfer, olisah2024convolutional} often compound complexity without addressing deployment barriers.

\subsection{Multi-Attribute Prediction and Explainability}

Ripeness and firmness are biologically coupled through ethylene production, cell wall breakdown, and chlorophyll degradation \citep{feng2023nondestructive, varga2021measuring}, yet most research treats them as independent prediction problems. \citet{feng2023nondestructive} found advantages for joint prediction over independent models in loquat quality assessment, and multi-modal approaches combining hyperspectral and RGB data show 10--15\% accuracy improvements \citep{garillos2021multimodal}. Yet single-task evaluation persists despite the fact that isolated quality attributes rarely drive commercial sorting decisions. This study adopts a paired single-task evaluation framework assessing ripeness and firmness jointly via aggregated metrics, while training separate models per task.

SHAP and LIME analyses identify biologically meaningful wavelengths: 680\,nm for chlorophyll absorption, 760\,nm for red-edge transitions, 970\,nm for water content \citep{ahmed2024advancing}. These findings align with plant physiology knowledge. If only 6--15 wavelengths contain most discriminative information \citep{nagasubramanian2018genetic, ahmed2024reconstruction}, full-spectrum systems may be overengineered and targeted multispectral sensors could achieve comparable performance at reduced cost.

\subsection{Benchmarking Gaps and Deployment Challenges}

Most studies evaluate 2--5 algorithms under carefully selected conditions \citep{yang2025hyperspectral}. Reported accuracies range from 73\% to 99\% for similar classification tasks \citep{vignati2023freshcut}. The DeepHS Fruit dataset \citep{varga2021measuring} provides standardized data across five fruit varieties (avocado, kiwi, mango, kaki/persimmon, papaya) with ripeness and firmness annotations. \citet{benjmaa2025fruit} introduced Fruit-HSNet, reporting 70.73\% overall accuracy on this benchmark. Systematic evaluation of traditional machine learning methods on this benchmark remains limited, with most comparative studies focusing on deep learning architectures.

Hyperspectral systems cost \$10{,}000--100{,}000 compared to \$500--5{,}000 for RGB alternatives, require specialized operators, and experience 15--30\% accuracy degradation under variable field conditions \citep{min2023decay}. Processing speed requirements for sorting applications (under 100 ms per sample) eliminate most deep learning approaches. RGB-reconstructed hyperspectral images \citep{ahmed2024reconstruction} and edge computing solutions \citep{lanke2025trends} address symptoms rather than causes.

No existing study systematically benchmarks a broad set of traditional ML algorithms on multi-attribute fruit quality prediction with controlled preprocessing ablation.

\section{Materials and Methods}
\label{sec:methods}

\subsection{Dataset Description}

We used the DeepHS Fruit dataset (version 2), a publicly available hyperspectral imaging dataset for fruit quality assessment research \citep{varga2021measuring}. The dataset is hosted on GitHub\footnote{https://github.com/cogsys-tuebingen/deephs\_fruit} and provides hyperspectral data with quality annotations for benchmarking. Version~2 corrects a potential test-set contamination issue in the original split by ensuring that balanced training variants do not alter the benchmark test partition; train, validation, and test sets are validated for zero sample-level overlap.

We use the dataset-provided VIS benchmark split as our fixed evaluation setting (unbalanced train/test: train $n$=381, test $n$=138; total $n$=519). Fruit-type distribution in this split is avocado ($n$=170, 32.8\%), kiwi ($n$=162, 31.2\%), mango ($n$=68, 13.1\%), kaki/persimmon ($n$=68, 13.1\%), and papaya ($n$=51, 9.8\%). The VIS subset spans 47 acquisition days, providing temporal variation in fruit maturation states. For the stratified resplit condition, we keep the benchmark test set fixed but replace the training set with a fruit-balanced training set ($n$=414; avocado $n$=130, kiwi $n$=130, mango $n$=56, kaki/persimmon $n$=55, papaya $n$=43), improving balance without breaking comparability.

The full DeepHS Fruit dataset includes multiple hyperspectral imaging systems (Specim FX10, INNO-SPEC Redeye 1.7, and Corning microHSI 410 Vis-NIR). To maintain spectral consistency and comparability with existing benchmarks, this study restricts evaluation to the VIS camera (Specim FX10, 398--1004\,nm), and all reported results reflect this sensor and its wavelength range. The FX10 covers only the short-NIR region; true NIR water absorption bands (1200--1450\,nm) that are commonly used in firmness assessment are absent from these data.

Each sample includes quality annotations from destructive measurements. Ripeness classification follows a three-class scheme based on visual assessment and destructive sampling \citep{varga2021measuring}; inter-annotator agreement was not reported in the original dataset publication. Classes are: unripe ($n$=116, 22.4\%), perfect ($n$=273, 52.6\%), and overripe ($n$=130, 25.0\%). Perfect ripeness samples are overrepresented at 52.6\%, while unripe and overripe samples are more evenly distributed.

Firmness measurements (grams-force from penetrometer testing) are categorized into three classes derived from cross-fruit distribution analysis: soft (0--1000\,gf, $n$=220, 42.4\%), medium (1001--2500\,gf, $n$=186, 35.8\%), and firm (2501+\,gf, $n$=99, 19.1\%). The distribution skews toward softer fruits (42.4\% soft vs 19.1\% firm). Samples with missing firmness measurements ($n$=14, 2.7\%; train $n$=13, test $n$=1) are retained and encoded as a separate ``Unknown'' label during training to avoid discarding data; they are excluded from all quality-metric computations reported in the tables.

\begin{figure*}[htbp!]
\centering
\includegraphics[width=\textwidth]{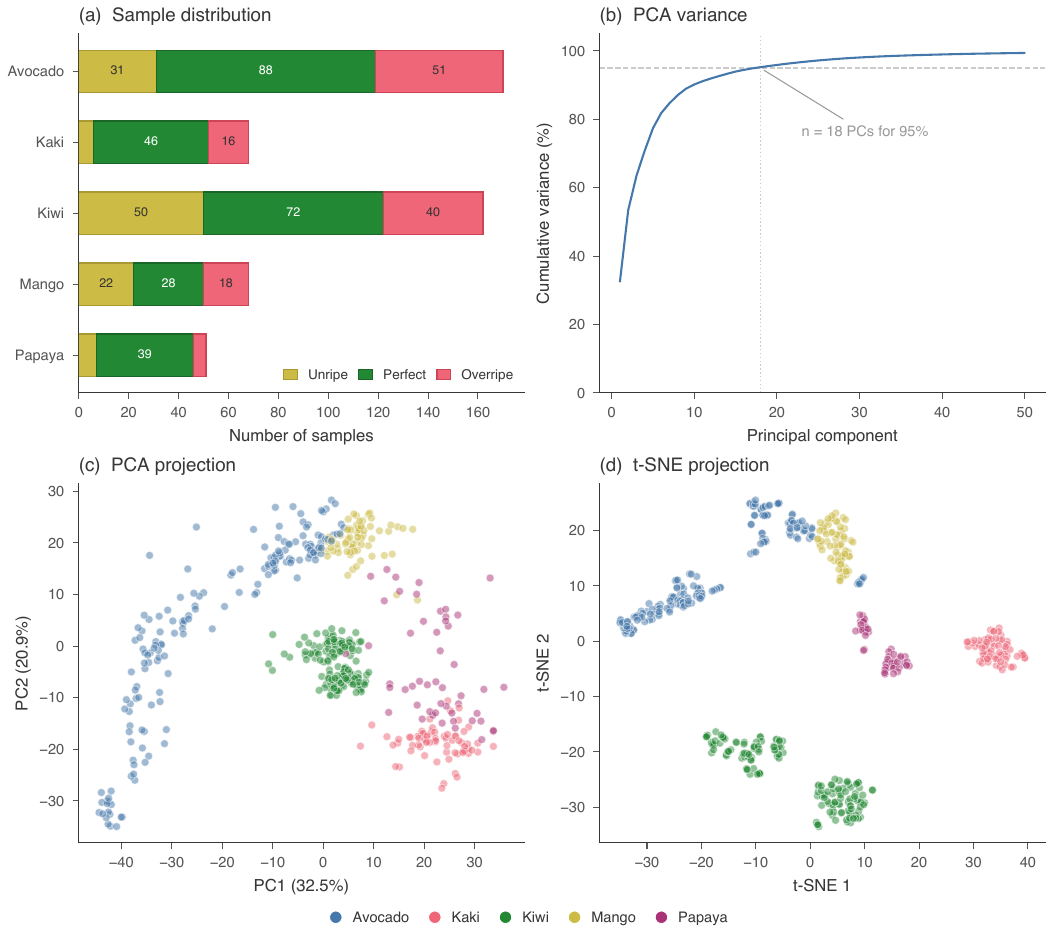}
\caption{Dataset overview. \textbf{(a)} Sample distribution across five fruit types and three ripeness classes. \textbf{(b)} PCA cumulative variance: 18 components capture 95\% of variance in the 1{,}120-dimensional feature space. \textbf{(c)} PCA projection and \textbf{(d)} t-SNE projection colored by fruit type show species-driven cluster structure rather than quality-state separation.}
\label{fig:dataset}
\end{figure*}

Sample distribution shows fruit-type imbalances (avocado $n$=170 vs papaya $n$=51) and ripeness-class imbalances (perfect 52.6\% vs unripe 22.4\%, Fig.~\ref{fig:dataset}a). PCA analysis indicates that 95\% cumulative explained variance is achieved with approximately 18--20 principal components from the 1{,}120-dimensional feature space (Fig.~\ref{fig:dataset}b). Projection onto the first two principal components (Fig.~\ref{fig:dataset}c) reveals that cluster structure is driven primarily by fruit type rather than quality state. This species-dominated variance structure motivates per-task rather than joint classification approaches.

\subsection{Data Preprocessing and Feature Engineering}

\subsubsection{Spectral Feature Extraction}

The DeepHS Fruit dataset provides hyperspectral data from three imaging systems (Specim FX10, INNO-SPEC Redeye 1.7, Corning microHSI 410 Vis-NIR). We use only the VIS camera (Specim FX10; 224 bands, 398--1004\,nm) for spectral consistency. We consume the dataset-provided extracted spectral feature tables (derived from background-masked hyperspectral cubes to isolate fruit pixels) and do not re-run pixel-level segmentation within the benchmarking pipeline.

An average spectrum is calculated for each sample by averaging spectral values across all foreground (fruit) pixels for each wavelength band:
\begin{equation}
\bar{S}(\lambda) = \frac{1}{|\mathcal{F}|} \sum_{(i,j) \in \mathcal{F}} S_{i,j}(\lambda)
\end{equation}
where \( S_{i,j}(\lambda) \) represents the spectral value at wavelength \( \lambda \) for pixel \((i, j)\), and \( \mathcal{F} \) denotes the set of foreground pixels. Each hyperspectral cube reduces to a 224-dimensional spectral signature per sample.

\subsubsection{Derived Spectral Features}

The combined feature set concatenates five spectral representations from the extracted feature tables, totaling 1{,}120 dimensions (5 representations $\times$ 224 spectral bands).

The average spectrum $\bar{S}(\lambda)$ captures absolute spectral reflectance across the VIS--NIR range. The first derivative highlights regions of rapid spectral change, emphasizing transitions related to chlorophyll degradation, carotenoid accumulation, and other biochemical changes:
\begin{equation}
D_1(\lambda) = \frac{d\bar{S}(\lambda)}{d\lambda}.
\end{equation}

Continuum removal normalizes the spectrum by fitting a convex hull and dividing the original spectrum by this hull, highlighting absorption bands such as chlorophyll absorption around 680\,nm:
\begin{equation}
CR(\lambda) = \frac{\bar{S}(\lambda)}{C(\lambda)},
\end{equation}
where $C(\lambda)$ is the continuum value at wavelength $\lambda$.

The standard normal variate (SNV) standardizes each mean spectrum:
\begin{equation}
SNV(\lambda) = \frac{\bar{S}(\lambda) - \mu}{\sigma},
\end{equation}
where $\mu$ and $\sigma$ are the mean and standard deviation of $\bar{S}(\lambda)$. Applied to per-sample mean spectra (averaged across pixels), SNV corrects for baseline offset and gain differences between samples; it does not correct within-pixel multiplicative scatter in the sense used for raw reflectance cubes.

Finally, the first derivative of the continuum-removed spectrum combines absorption-emphasizing and edge-enhancing properties:
\begin{equation}
D_1^{CR}(\lambda) = \frac{dCR(\lambda)}{d\lambda}.
\end{equation}

These five representations capture complementary aspects of spectral variation.

\begin{figure*}[htbp!]
\centering
\includegraphics[width=\textwidth]{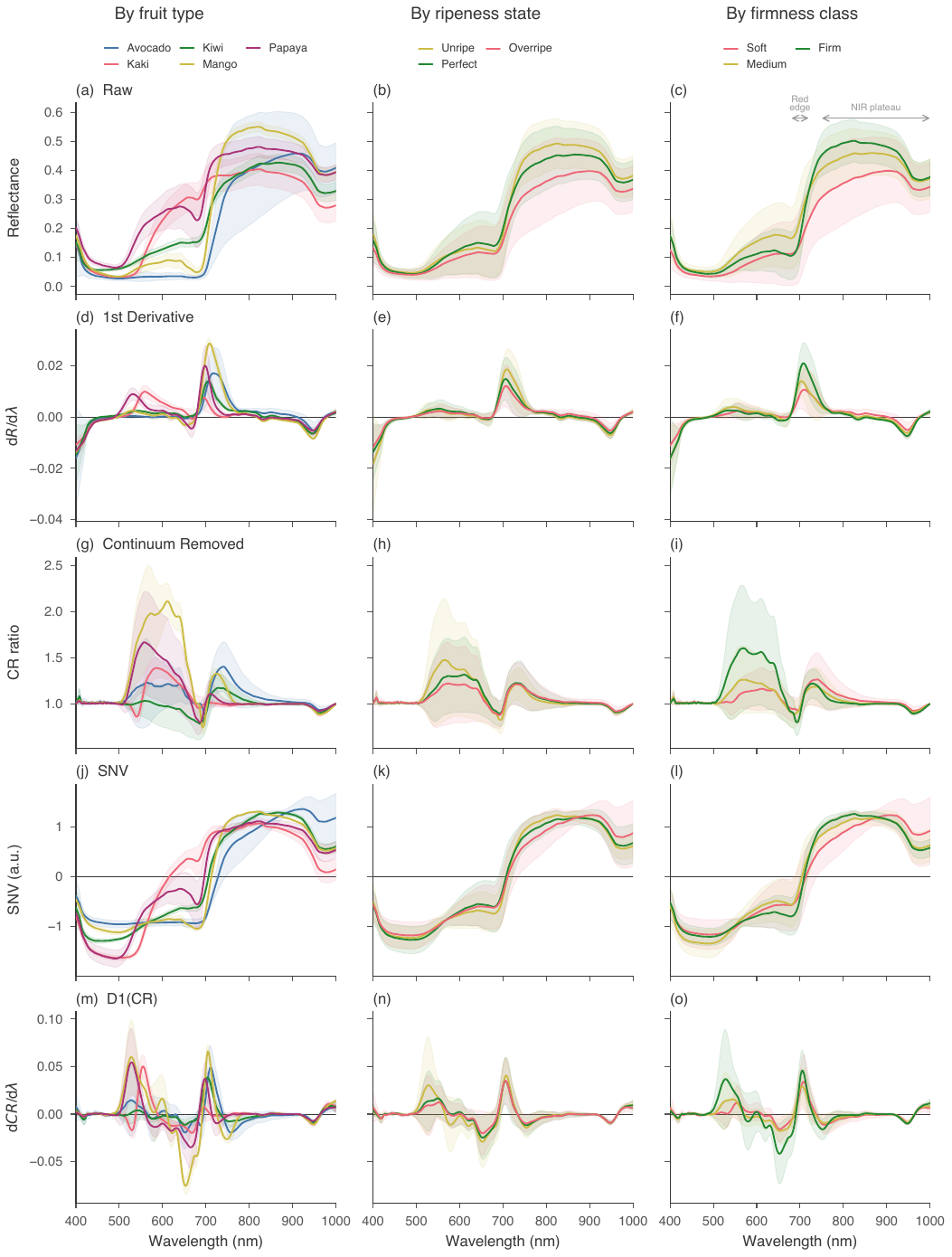}
\caption{Five spectral preprocessing types (rows) across the 224 bands (400--1000\,nm), grouped by fruit type, ripeness state, and firmness class (columns). Shaded bands: mean $\pm$ std. Rows: (a--c)~raw, (d--f)~1st derivative, (g--i)~continuum removed, (j--l)~SNV, (m--o)~D1(CR). Inter-species variation dominates; class separation is clearest in derivative features near the red edge ($\sim$700\,nm).}
\label{fig:preprocessing-types}
\end{figure*}

Raw spectral reflectance signatures for the five fruit types exhibit characteristic pigment-related variations across ripeness states, particularly in chlorophyll absorption regions ($\sim$680\,nm) and red-edge transitions ($\sim$700--750\,nm). Near-infrared features (700--1000\,nm) show structure-related variations correlated with firmness. These signatures arise from changes in cellular architecture and water content during maturation (Fig.~\ref{fig:preprocessing-types}).

\subsubsection{Dataset Balancing and Train/Test Splits}

The original DeepHS Fruit splits show class imbalances for certain fruit types, ripeness states, and firmness categories. We distinguish two categories of data balancing:

\textit{Training set replacement.} The stratified resplit condition replaces the training partition with a fruit-balanced set ($n$=414 vs the original $n$=381), keeping the benchmark test set fixed. This changes both the sample composition and size (an 8.7\% increase in training samples), so its advantage may partly reflect additional data rather than purely improved balance.

\textit{Training set augmentation.} SMOTE, random oversampling, and random undersampling modify the class distribution within the original training partition without changing which real samples are included. Sampling is applied independently per task (ripeness or firmness); the test partition is kept fixed.

Firmness is discretized into three bins: soft (0--1000\,gf), medium (1001--2500\,gf), and firm (2501+\,gf), derived from cross-fruit firmness distribution analysis. The original DeepHS dataset defines fruit-specific thresholds (e.g., $<$900/900--1200/$>$1200\,g/cm$^2$ for avocado; \citealt{varga2021measuring}), but universal bins enable direct cross-fruit comparison across the full firmness spectrum.

The original unbalanced split is preserved for comparability with benchmarks from Varga et al. \citep{varga2021measuring} and Ben Jmaa et al. \citep{benjmaa2025fruit}.

\subsubsection{Additional Preprocessing Strategies}

When enabled, PCA reduces dimensionality while retaining 95\% of variance, compressing the 1{,}120-dimensional feature space to approximately 18--20 principal components. PCA is integrated within the scikit-learn Pipeline class: the transformation is fit only on training data and applied to test data, and in Phase 4 cross-validation PCA is fit separately within each fold. The five spectral representations are concatenated before PCA, so it operates on the full combined feature space.

SMOTE (Synthetic Minority Oversampling Technique) generates synthetic samples for minority classes using the imbalanced-learn\footnote{imbalanced-learn: \url{https://imbalanced-learn.org}} library ($k$=5 nearest neighbors), applied independently per task after train/test splitting but before model training. For the firmness task, which includes 14 samples (2.7\%) with missing values encoded as a separate class, SMOTE treats this as a fourth firmness category. Random oversampling duplicates minority class samples; random undersampling removes majority class samples.

Phase 2 evaluates these strategies across 200 configurations (5 data balancing strategies $\times$ 2 PCA options $\times$ 20 models).

\subsubsection{Wavelength Subset Experiments (VIS-3 and RGB)}

Two three-band configurations are evaluated to assess the necessity of full-spectrum sensing.

\textit{VIS-3 (XAI-derived).} Three visible-range bands are selected from hyperspectral data at indices \(\{18, 52, 89\}\), corresponding to wavelengths \(\{448, 540, 640\}\,\)nm. These bands were identified through consensus feature ranking from ExtraTrees explainability analysis (Section~\ref{sec:results}), selecting the top joint-ranked visible-range bands ($<$700\,nm) across both ripeness and firmness tasks. We term this the ``VIS-3 subset'' to emphasize that these bands are selected from hyperspectral cubes, not acquired by an independent RGB sensor.

\textit{RGB (approximate).} Three bands at indices \(\{19, 56, 93\}\), corresponding to wavelengths \(\{450, 550, 651\}\,\)nm, approximate the center wavelengths of standard Bayer-pattern RGB sensors. This configuration tests whether XAI-guided band selection outperforms generic RGB-center wavelengths, thereby isolating the contribution of precise band placement from the effect of spectral dimensionality reduction.

Both configurations apply the same five spectral transformations, resulting in a 15-dimensional feature vector (3 bands \(\times\) 5 methods) compared to the 1{,}120-dimensional full-spectrum vector. Derivatives and continuum removal of three non-contiguous bands lack the physical meaning they carry over continuous spectra; reduced-band results therefore represent an upper bound on what a true 3-band sensor could achieve, not a realistic sensor comparison. Both pipelines follow identical experimental procedures to the full-spectrum pipeline.

\subsection{Experimental Design}

The six-phase methodology trains each algorithm separately for ripeness and firmness, with performance reported jointly through aggregated metrics.

\subsubsection{Pipeline Phases}

Each model is trained separately for ripeness and firmness prediction, with performance evaluated using overall accuracy (OA, Eq.~\ref{eq:oa}). We term this ``paired single-task evaluation'' to distinguish it from multi-task learning architectures that share representations across tasks. Each task is modeled independently; only the evaluation metric aggregates performance. We use separate models rather than joint multi-output classification for four reasons: (1)~many evaluated algorithms do not natively support multi-output classification; (2)~separate models allow independent hyperparameter tuning (firmness accuracy exceeds ripeness accuracy by approximately 25~pp, Table~\ref{tab:task-asymmetry}); (3)~agricultural operations may prioritize one attribute depending on market demands; and (4)~existing literature predominantly uses single-task evaluation. This approach does not exploit shared representations that joint models might capture; future work should compare separate-task versus joint multi-task architectures.

Phase~1 establishes baseline performance for 20 algorithms on the original imbalanced dataset using default hyperparameters, evaluated using accuracy, F1-macro, and F1-weighted metrics for each task. Phase~2 evaluates preprocessing strategies using a full-factorial design across data balancing strategy (five levels: original imbalanced, stratified resplit, SMOTE, random oversampling, random undersampling), PCA dimensionality reduction (enabled at 95\% variance or disabled), and model selection (20 algorithms), generating 200 configurations. The best Phase~2 preprocessing setting per modality is selected for Phase~3 based on overall accuracy. This selection uses test-set performance, introducing indirect information leakage: the chosen preprocessing was optimized, in part, on the same held-out set used for final evaluation. Phase~4 cross-validation provides a partial mitigation, but the preprocessing choice itself was not cross-validated.

Phase~3 applies Bayesian optimization via Optuna\footnote{Optuna: \url{https://optuna.org}} \citep{akiba2019optuna} with the best Phase~2 preprocessing configuration. For each model, 100 trials using Tree-structured Parzen Estimator (TPE) sampling \citep{bergstra2011algorithms} maximize $\overline{\mathrm{F1}}_{\mathrm{macro}}$ (Eq.~\ref{eq:f1macro}).
Search spaces are model-specific (Table~\ref{tab:model-overview}): tree-based models optimize depth, minimum samples per leaf, and ensemble size; linear models optimize regularization; neural networks optimize learning rate, hidden layer sizes, and activations. Optimization uses stratified 70/30 train/validation splits, with final models retrained on the full training set. Total: 2{,}000 trials per modality (20 models $\times$ 100 trials), 4{,}000 trials across both modalities.

Phase~4 validates model performance through 10-fold stratified cross-validation using optimal hyperparameters from Phase~3 and best preprocessing from Phase~2. Results are reported as mean $\pm$ standard deviation across folds, with 95\% confidence intervals and coefficient of variation. Phases~1--3 provide exploratory results on a single fixed split; Phase~4 cross-validation is the authoritative basis for algorithm ranking and all primary conclusions.

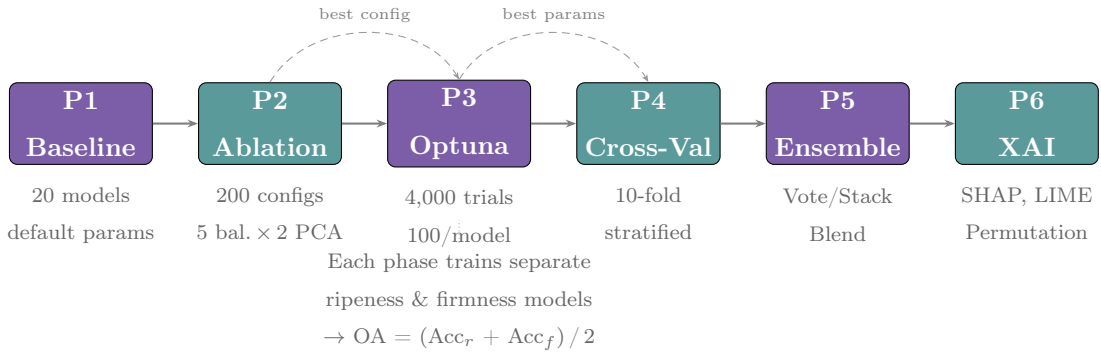
\begin{figure*}[htbp!]
\centering
\begin{tikzpicture}[
    node distance=0.4cm and 0.6cm,
    phase/.style={draw, rounded corners=3pt, minimum height=0.9cm, minimum width=1.9cm,
                  font=\footnotesize\bfseries, text=white, align=center, inner sep=3pt},
    annot/.style={font=\scriptsize, text=black!60, align=center},
    arr/.style={-{Stealth[length=4pt]}, thick, black!50},
    feed/.style={-{Stealth[length=3pt]}, densely dashed, thin, black!40},
]
% Phase boxes — alternating purple/teal
\definecolor{cpurple}{HTML}{7B5EA7}
\definecolor{cteal}{HTML}{5A9A9B}

\node[phase, fill=cpurple]                 (p1) {P1\\Baseline};
\node[phase, fill=cteal, right=of p1]      (p2) {P2\\Ablation};
\node[phase, fill=cpurple, right=of p2]    (p3) {P3\\Optuna};
\node[phase, fill=cteal, right=of p3]      (p4) {P4\\Cross-Val};
\node[phase, fill=cpurple, right=of p4]    (p5) {P5\\Ensemble};
\node[phase, fill=cteal, right=of p5]      (p6) {P6\\XAI};

% Annotations below
\node[annot, below=0.15cm of p1] {20 models\\default params};
\node[annot, below=0.15cm of p2] {200 configs\\5 bal.\,$\times$\,2 PCA};
\node[annot, below=0.15cm of p3] {4{,}000 trials\\100/model};
\node[annot, below=0.15cm of p4] {10-fold\\stratified};
\node[annot, below=0.15cm of p5] {Vote/Stack\\Blend};
\node[annot, below=0.15cm of p6] {SHAP, LIME\\Permutation};

% Forward arrows
\draw[arr] (p1) -- (p2);
\draw[arr] (p2) -- (p3);
\draw[arr] (p3) -- (p4);
\draw[arr] (p4) -- (p5);
\draw[arr] (p5) -- (p6);

% Feedback arrows
\draw[feed] (p2.north) to[out=60,in=120] node[above, font=\tiny, text=black!50] {best config} (p3.north);
\draw[feed] (p3.north) to[out=60,in=120] node[above, font=\tiny, text=black!50] {best params} (p4.north);

% Paired single-task annotation
\node[annot, below=1.0cm of p3, text width=4cm] (dual) {Each phase trains separate\\ripeness \& firmness models\\$\rightarrow$ OA = (Acc$_r$ + Acc$_f$)\,/\,2};
\draw[densely dotted, thin, black!30] (dual.north) -- ++(0,0.3);
\end{tikzpicture}
\caption{Six-phase benchmark pipeline. Each phase trains independent ripeness and firmness models, combined via overall accuracy (OA). Dashed arrows indicate configuration propagation between phases.}
\label{fig:pipeline}
\end{figure*}

Phase~5 combines top models from Phase~4 using soft voting, hard voting, stacking, and blending. Phase~6 performs dedicated explainability analysis for the best-performing single model using SHAP\footnote{SHAP: \url{https://github.com/shap/shap}} \citep{lundberg2017unified}, LIME\footnote{LIME: \url{https://github.com/marcotcr/lime}} \citep{ribeiro2016should}, and permutation importance \citep{breiman2001random}, producing wavelength-level interpretations for both tasks. The full pipeline architecture is illustrated in Fig.~\ref{fig:pipeline}.

\subsubsection{Benchmark Reference}

We reference Fruit-HSNet's reported 70.73\% overall accuracy \citep{benjmaa2025fruit} as a contextual data point, not a controlled comparison. Fruit-HSNet processes full 64$\times$64 pixel hyperspectral cubes through a dual-branch network that extracts Fourier-domain spectral features alongside central-pixel spatial signatures, trained with GPU acceleration, focal loss, and spatial data augmentation including random cropping, flipping, rotation, and noise injection. Our approach discards spatial information entirely, representing each sample by its per-pixel-averaged mean spectrum (a 224-dimensional vector retaining no spatial structure) processed on a consumer CPU in under one second. The following protocol differences preclude direct comparison: (1)~Fruit-HSNet used a different train/test partition, (2)~label preprocessing and class definitions may differ, and (3)~Fruit-HSNet is a deep learning architecture while this study evaluates the scikit-learn ecosystem. Our Phase~3 result (ExtraTrees, 75.0\%) carries a Wilson 95\% confidence interval of [67.6\%, 81.8\%], which contains Fruit-HSNet's reported value, and no statistically meaningful difference can be claimed.

Phase 4 cross-validation provides statistical estimates through 10-fold stratified sampling (mean, standard deviation, coefficient of variation, 95\% confidence intervals). We apply the Friedman test for omnibus comparison of all 20 classifiers on per-fold F1 scores, followed by Nemenyi post-hoc tests with family-wise error control (Table~\ref{tab:model-ranks}). Cohen's $d$ effect sizes quantify task asymmetry (Table~\ref{tab:stat-tests}). Phase~4 cross-validated rankings are the authoritative ranking for algorithm comparison; Phase~3 single-split results are retained only for protocol-matched reference against Fruit-HSNet.
%\subsubsection{Reproducibility}
Reproducibility details are provided in Table~\ref{tab:hardware}.

\subsection{Machine Learning Models}

We evaluate 20 classical and gradient-boosted machine learning algorithms from seven algorithmic families, including PLS-DA (Partial Least Squares Discriminant Analysis), the standard chemometric baseline for spectral classification. PLS-DA is implemented as a custom scikit-learn wrapper around \texttt{PLSRegression} with dummy-coded class labels and argmax prediction; \texttt{n\_components} is clamped at fit time to $\min(n_\mathrm{samples}, n_\mathrm{features}, n_\mathrm{classes})$. Note that the sklearn \texttt{MLPClassifier} used here is a shallow multi-layer perceptron; it shares an API with other scikit-learn classifiers but is not a deep learning model. Models are implemented using scikit-learn\footnote{scikit-learn: \url{https://scikit-learn.org}} \citep{pedregosa2011scikit}, XGBoost\footnote{XGBoost: \url{https://xgboost.ai}} \citep{chen2016xgboost}, and LightGBM\footnote{LightGBM: \url{https://lightgbm.readthedocs.io}} \citep{ke2017lightgbm}. Each model is integrated into a pipeline with feature scaling (StandardScaler) and optional PCA. Table~\ref{tab:model-overview} provides a structured overview of all models, including abbreviations, family membership, hyperparameter counts, Optuna search spaces, and native feature importance support.

All models use fixed random state (seed=42), early stopping where applicable, and class balancing for imbalanced datasets. A complete overview of all models with Optuna search spaces is provided in Table~\ref{tab:model-overview} (Appendix); best hyperparameters from Bayesian optimization are listed in Table~\ref{tab:best-hyperparams}.

\subsection{Evaluation Metrics}

We report per-task accuracies for ripeness ($r$) and firmness ($f$), overall accuracy (OA) defined as the arithmetic mean of task accuracies, and F1-scores (both macro-averaged and weighted). Per-class metrics for class $c$:
\begin{equation}
\mathrm{Precision}_c = \frac{TP_c}{TP_c + FP_c},
\end{equation}
\begin{equation}
\mathrm{Recall}_c = \frac{TP_c}{TP_c + FN_c},
\end{equation}
\begin{equation}
\mathrm{F1}_c = \frac{2\,\mathrm{Precision}_c\,\mathrm{Recall}_c}{\mathrm{Precision}_c + \mathrm{Recall}_c}.
\end{equation}
Task-level aggregates for task $t \in \{r, f\}$ with class set $C_t$:
\begin{equation}
\mathrm{F1}_{\mathrm{macro}}^{(t)} = \frac{1}{|C_t|}\sum_{c \in C_t}\mathrm{F1}_c,
\end{equation}
\begin{equation}
\mathrm{F1}_{\mathrm{wt}}^{(t)} = \sum_{c \in C_t}\frac{n_c}{N_t}\,\mathrm{F1}_c.
\end{equation}
Overall metrics average across both tasks:
\begin{equation}
\label{eq:oa}
\mathrm{OA} = \frac{\mathrm{Acc}^{(r)} + \mathrm{Acc}^{(f)}}{2},
\end{equation}
\begin{equation}
\label{eq:f1macro}
\overline{\mathrm{F1}}_{\mathrm{macro}} = \frac{\mathrm{F1}_{\mathrm{macro}}^{(r)} + \mathrm{F1}_{\mathrm{macro}}^{(f)}}{2}.
\end{equation}

Equal weighting treats both quality attributes as equally relevant for holistic assessment; deployment scenarios prioritizing one attribute can reweight accordingly. OA is the primary benchmark comparison metric; $\overline{\mathrm{F1}}_{\mathrm{macro}}$ is the optimization objective for Phase~3 and Phase~4 cross-validation. Cross-validation reports mean $\pm$ standard deviation, coefficient of variation, and 95\% confidence intervals. Efficiency metrics include training time (s), inference time (s), and model size (MB).

\subsection{Computational Environment}

All experiments were conducted on consumer-grade hardware, confirming feasibility on consumer hardware without specialized infrastructure (Table~\ref{tab:hardware}). Hardware configuration consists of MacBook Air M4 (2024) with Apple Silicon M4 chip, 24\,GB unified memory, SSD storage, and macOS (arm64). Software stack includes Python 3.13, scikit-learn 1.7.1, XGBoost 3.0.4, LightGBM 4.6.0, Optuna 4.5.0, and imbalanced-learn 0.14.0.

CO$_2$ emissions were negligible across all models ($<$0.1\,g per model on CPU-only consumer hardware; estimated via CodeCarbon\footnote{CodeCarbon: \url{https://codecarbon.io}} methodology \citep{schmidt2021codecarbon}).

\section{Results}
\label{sec:results}

Evaluation spans 20 algorithms across three modalities (full-spectrum: 1{,}120 features; VIS-3 subset: 15 features; RGB approximate: 15 features) and six experimental phases, including 10-fold cross-validation and 4{,}000 hyperparameter-optimization trials. A Friedman test on per-fold overall F1 scores confirmed significant differences among classifiers ($\chi^2(19) = 144.8$, $p < 10^{-20}$); the Nemenyi post-hoc critical difference diagram (Fig.~\ref{fig:cd-diagram} in Appendix, Table~\ref{tab:model-ranks}) identifies 46 significantly different pairs. %The performance of the models is summarized in Table~\ref{tab:comprehensive}.

\subsection{Pipeline Performance}

\begin{table*}[t]
\centering
\caption{Model performance and efficiency across pipeline phases (full-spectrum, 1{,}120 features). Models sorted by Phase~4 cross-validated overall accuracy.\protect\footnotemark\ P4 shows 10-fold stratified cross-validation accuracy (mean $\pm$ std). P3 Wilson 95\% CI (binomial, $n$=138) shown in brackets. \textbf{Bold} = best in column; \ul{underline} = second-best. Arrows indicate preferred direction.}
\label{tab:comprehensive}
\small
\setlength{\tabcolsep}{4pt}
\resizebox{\textwidth}{!}{%
\begin{tabular}{clllllll lr r}
\toprule
Rk & Model & Family & P1$\uparrow$ & P2$\uparrow$ & P2 Config & P3$\uparrow$ & P3 CI & P4 CV (\%)$\uparrow$ & Time$\downarrow$ & Size$\downarrow$ \\
   &       &        & OA           & OA           &           & OA           & [95\%] &                      & (s)              & (MB) \\
\midrule
  1 & XGBoost & Boosting & 67.4 & 67.4 & SMOTE / No PCA & 65.6 & [57.0, 72.7] & \textbf{82.3} $\pm$ 5.4 & 5.9 & 0.72 \\
  2 & HistGradientBoosting & Boosting & 65.2 & 68.1 & SMOTE / No PCA & 67.4 & [59.2, 74.6] & \ul{82.2} $\pm$ 5.4 & 4.6 & 5.05 \\
  3 & LGBM & Boosting & 68.1 & 69.2 & Strat. Resplit / No PCA & 69.2 & [61.4, 76.6] & \ul{82.2} $\pm$ 4.9 & 4.7 & 1.64 \\
  4 & ExtraTrees & Tree-based & \textbf{71.7} & \textbf{75.0} & Strat. Resplit / No PCA & \textbf{75.0} & [67.6, 81.8] & \ul{82.2} $\pm$ 4.7 & \textbf{0.2} & 2.77 \\
  5 & SVC\_RBF & SVM & 56.2 & 57.6 & SMOTE / No PCA & 56.9 & [48.2, 64.5] & 81.1 $\pm$ 4.0 & 0.6 & 5.16 \\
  6 & GradientBoosting & Boosting & 64.1 & 67.4 & Oversample / No PCA & 63.8 & [55.5, 71.3] & 80.6 $\pm$ 5.8 & 18.8 & 0.41 \\
  7 & RandomForest & Tree-based & \ul{69.2} & \ul{73.6} & Strat. Resplit / No PCA & \ul{73.6} & [66.0, 80.5] & 79.8 $\pm$ 4.8 & 0.5 & 1.44 \\
  8 & KNeighbors & Other & 66.3 & 66.3 & Strat. Resplit / PCA & 65.2 & [57.0, 72.7] & 78.9 $\pm$ 4.4 & \textbf{0.2} & 7.14 \\
  9 & LogisticRegression & Linear & 50.4 & 51.4 & Oversample / No PCA & 46.7 & [38.3, 54.7] & 78.0 $\pm$ 5.2 & 0.9 & 0.11 \\
  10 & SVC\_Linear & SVM & 53.3 & 53.3 & Original / No PCA & 49.6 & [41.1, 57.5] & 76.9 $\pm$ 5.2 & \ul{0.4} & 3.64 \\
  11 & MLPClassifier & Other & 61.2 & 61.2 & Original / No PCA & 56.2 & [48.2, 64.5] & 75.6 $\pm$ 6.0 & \textbf{0.2} & 5.22 \\
  12 & LDA & Linear & 35.5 & 58.7 & Strat. Resplit / PCA & 38.4 & [30.7, 46.7] & 74.7 $\pm$ 4.6 & \textbf{0.2} & 0.23 \\
  13 & SGDClassifier & Linear & 58.0 & 58.7 & Strat. Resplit / PCA & 58.3 & [49.6, 65.9] & 73.4 $\pm$ 4.2 & \textbf{0.2} & 0.11 \\
  14 & DecisionTree & Tree-based & 49.3 & 65.2 & SMOTE / No PCA & 58.7 & [50.4, 66.6] & 72.8 $\pm$ 5.2 & \ul{0.4} & 0.07 \\
  15 & Ridge & Linear & 42.4 & 59.4 & Strat. Resplit / PCA & 46.0 & [38.3, 54.7] & 72.8 $\pm$ 3.9 & \textbf{0.2} & 0.11 \\
  16 & AdaBoost & Boosting & 57.2 & 62.7 & Original / PCA & 62.0 & [54.0, 70.0] & 69.3 $\pm$ 5.9 & 2.5 & 0.12 \\
  17 & PassiveAggressive & Linear & 49.6 & 58.7 & SMOTE / PCA & 54.0 & [45.3, 61.7] & 68.8 $\pm$ 5.4 & \textbf{0.2} & 0.11 \\
  18 & PLSDA & Chemometric & 61.6 & 61.6 & Original / No PCA & 59.1 & [51.1, 67.3] & 57.8 $\pm$ 4.5 & \textbf{0.2} & 0.42 \\
  19 & BernoulliNB & Naive Bayes & 59.1 & 59.1 & Undersample / No PCA & 56.9 & [48.2, 64.5] & 54.9 $\pm$ 5.5 & \textbf{0.2} & 0.17 \\
  20 & GaussianNB & Naive Bayes & 60.5 & 61.2 & SMOTE / PCA & 59.1 & [51.1, 67.3] & 52.8 $\pm$ 4.9 & \textbf{0.2} & 0.17 \\
\bottomrule
\end{tabular}%
}
\footnotetext{P3 and P4 rankings correlate at Spearman $\rho=0.51$ ($p = 0.020$); P4 cross-validated rankings are authoritative. P3 CIs are Wilson 95\% intervals ($n$=138). Protocol differences with Fruit-HSNet are documented in Section~\ref{sec:methods}. CO$_2$ emissions negligible ($<$0.1\,g per model); column omitted. The P2 Config column shows the per-model best Phase~2 configuration; Phase~3 optimization uses a single global best configuration (stratified resplit / no PCA) for all models. PLS-DA P4 CV metrics should be interpreted with caution: a pipeline anomaly produced identical per-fold ripeness and firmness accuracies, likely due to incorrect task separation during cross-validation (see Section~\ref{sec:discussion}).}
\end{table*}

\begin{table}[htbp!]
\centering
\caption{Model rankings from 10-fold cross-validation (Friedman/Nemenyi, $\alpha=0.05$). Rank = mean rank across folds (lower is better). Models connected by a bar in Fig.~\ref{fig:cd-diagram} are not significantly different.}
\label{tab:model-ranks}
\small
\begin{tabular}{@{}clccc@{}}
\toprule
Rank & Model & Avg.\ Rank$\downarrow$ & Mean F1$\uparrow$ & Family \\
\midrule
  1 & ET & \textbf{3.7} & \textbf{0.792 $\pm$ 0.077} & Tree-based \\
  2 & SVM-R & \ul{4.7} & \ul{0.787 $\pm$ 0.077} & SVM \\
  3 & KNN & 5.4 & 0.771 $\pm$ 0.061 & Other \\
  4 & GBC & 5.8 & 0.760 $\pm$ 0.064 & Boosting \\
  5 & RF & 6.2 & 0.758 $\pm$ 0.063 & Tree-based \\
  6 & LGBM & 6.2 & 0.755 $\pm$ 0.049 & Boosting \\
  7 & HGBC & 6.9 & 0.752 $\pm$ 0.048 & Boosting \\
  8 & SVM-L & 7.2 & 0.747 $\pm$ 0.067 & SVM \\
  9 & LR & 7.5 & 0.750 $\pm$ 0.077 & Linear \\
  10 & XGB & 7.6 & 0.743 $\pm$ 0.063 & Boosting \\
  11 & MLP & 7.7 & 0.725 $\pm$ 0.049 & Other \\
  12 & LDA & 12.2 & 0.671 $\pm$ 0.060 & Linear \\
  13 & DT & 13.0 & 0.659 $\pm$ 0.052 & Tree-based \\
  14 & SGD & 14.0 & 0.639 $\pm$ 0.055 & Linear \\
  15 & Ridge & 14.6 & 0.626 $\pm$ 0.041 & Linear \\
  16 & PA & 15.5 & 0.610 $\pm$ 0.057 & Linear \\
  17 & Ada & 15.7 & 0.609 $\pm$ 0.044 & Boosting \\
  18 & BNB & 18.1 & 0.526 $\pm$ 0.065 & Naive Bayes \\
  19 & PLS-DA & 18.7 & 0.491 $\pm$ 0.065 & Chemometric \\
  20 & GNB & 19.3 & 0.486 $\pm$ 0.049 & Naive Bayes \\
\bottomrule
\end{tabular}
\end{table}

\begin{table}[htbp!]
\centering
\caption{Summary of statistical tests and descriptive comparisons. Effect sizes follow Cohen's conventions: $|d|<0.2$ negligible, $0.2$--$0.5$ small, $0.5$--$0.8$ medium, $>0.8$ large. Friedman and Nemenyi tests are appropriate because models share the same CV folds. Firmness vs.\ ripeness comparisons involve the same 20 models trained on shared data; formal significance tests are not used because the observations are not independent. Phase~3 modality comparisons (HSI vs.\ VIS-3, HSI vs.\ RGB, VIS-3 vs.\ RGB) use paired Wilcoxon signed-rank tests on per-model overall accuracy.}
\label{tab:stat-tests}
\small
\setlength{\tabcolsep}{5pt}
\begin{tabular}{@{}lccc@{}}
\toprule
Comparison & Summary & $d$ & Effect \\
\midrule
Friedman (20 classifiers) & $\chi^2(19)=144.8$, $p < 10^{-20}$ & --- & --- \\
\quad Nemenyi pairs ($\alpha$=0.05) & 46 of 190 pairs differ & --- & --- \\
\midrule
Firmness $>$ Ripeness & 19/20 models; mean gap 25.0~pp & 1.84 & Large \\
\quad (IQR: 21.4--32.8~pp; exception: LDA) & & & \\
HSI vs.\ RGB & Cohen's $d$ from Phase~4 fold F1 & 1.17 & Large \\
HSI vs.\ VIS-3 & Phase~3 OA, 20 models; $p=0.46$ & 0.32 & Small \\
HSI vs.\ RGB & Phase~3 OA, 20 models; $p=0.39$ & 0.14 & Negl. \\
VIS-3 vs.\ RGB & Phase~3 OA, 20 models; $p=0.30$ & 0.27 & Small \\
\bottomrule
\end{tabular}
\end{table}

Table~\ref{tab:comprehensive} presents integrated results across all pipeline phases for the 20 evaluated models, sorted by Phase~4 cross-validated overall accuracy. XGBoost achieved the highest cross-validated OA ($82.3 \pm 5.4\%$), with HistGradientBoosting, LGBM, and ExtraTrees within 0.1 percentage points ($82.2\%$ each). In total, 11 of 20 models exceeded 75\% OA in 10-fold CV. ExtraTrees, which ranked first on the single benchmark split (Phase~3, 75.0\%), maintained strong cross-validated performance. Phase~4 cross-validated rankings are the authoritative comparison. Subsequent per-class and XAI analyses use ExtraTrees because it provides native feature importance scores and achieved the highest single-split accuracy on the benchmark partition used for Fruit-HSNet comparison. Training time was 0.2{}~seconds on consumer hardware.

\begin{figure*}[htbp!]
\centering
\includegraphics[width=\textwidth]{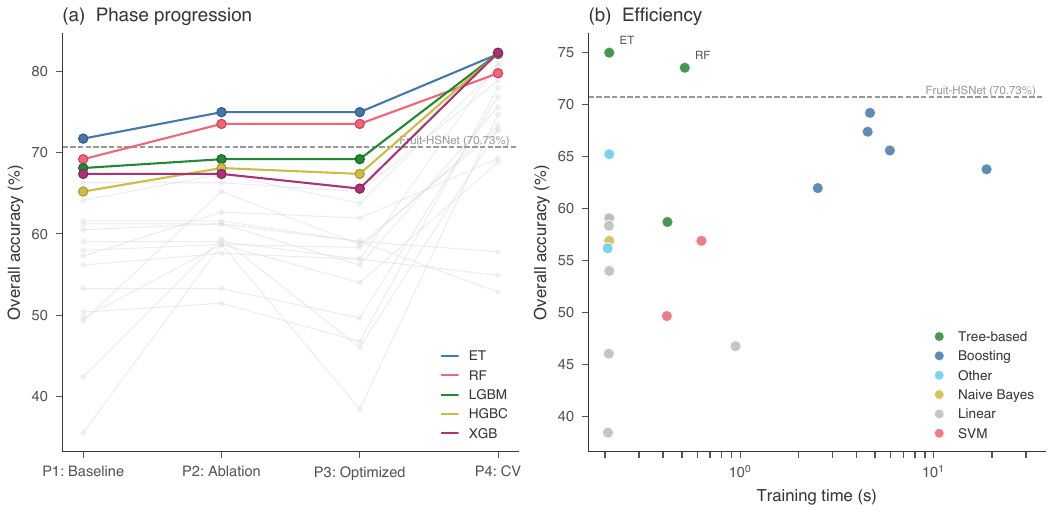}
\caption{Pipeline performance. (a)~Phase progression for all 20 models: connected dots trace each model's overall accuracy from baseline (P1) through ablation (P2, best config), optimization (P3), and 10-fold cross-validation (P4). Top-5 models highlighted; remaining in gray. The largest gains occur between P1 and P2, confirming that preprocessing contributes more than hyperparameter tuning for competitive models. (b)~Training-time efficiency (log scale) vs overall accuracy (Phase~3), colored by algorithmic family. Marker size reflects model storage size. Star markers denote Pareto-optimal models in model size vs accuracy space (6 models: ExtraTrees, RandomForest, XGBoost, GradientBoosting, AdaBoost, DecisionTree), representing the best size-accuracy tradeoff frontier. Tree-based methods achieve the highest accuracy with the shortest training times.}
\label{fig:pipeline-performance}
\end{figure*}

Preprocessing strategy substantially influenced performance for top models. Stratified resplit balancing improved ExtraTrees from 71.7\% (Phase~1 baseline) to 75.0\% (Phase~2), a 3.3 percentage point gain from data balancing alone. Phase~3 Bayesian optimization (100 trials per model) did not further improve ExtraTrees. For competitive models, preprocessing rather than hyperparameter tuning drove the performance gain (Fig.~\ref{fig:pipeline-performance}a). All 20 models trained in under 19~seconds on the consumer-grade hardware (MacBook Air M4). Inference latency (Phase~4) is well below the 100\,ms sorting-line threshold for all top models: ExtraTrees 2.1\,ms/sample, RandomForest 2.9\,ms/sample, LGBM 6.8\,ms/sample, XGBoost 34.1\,ms/sample (138 test samples). The most efficient model (ExtraTrees) is 90{}$\times$ faster than GradientBoosting while achieving 11.2 percentage points higher accuracy (Fig.~\ref{fig:pipeline-performance}b).

The Pareto frontier in model size vs.\ accuracy space (Fig.~\ref{fig:pipeline-performance}b, star markers) identifies six non-dominated models, namely ExtraTrees (75.0\%, 2.8~MB), RandomForest (73.6\%, 1.4~MB), XGBoost (65.6\%, 0.7~MB), GradientBoosting (63.8\%, 0.4~MB), AdaBoost (62.0\%, 0.1~MB), and DecisionTree (58.7\%, 0.07~MB). ExtraTrees achieves the highest accuracy at a storage footprint suitable for embedded sorting hardware, with inference latency of 2.1~ms per sample. The dominance of preprocessing over optimization across competitive models raises the question of which preprocessing decisions matter most and why.

\FloatBarrier
\subsection{Preprocessing Ablation}

Table~\ref{tab:ablation} summarizes the Phase~2 ablation study across 200 configurations (5 balancing strategies $\times$ 2 PCA options $\times$ 20 models). Stratified resplit achieved the highest mean overall accuracy (59.08\% without PCA), while random undersampling performed worst (46.56\%).

\begin{figure*}[htbp!]
\centering
\includegraphics[width=\textwidth]{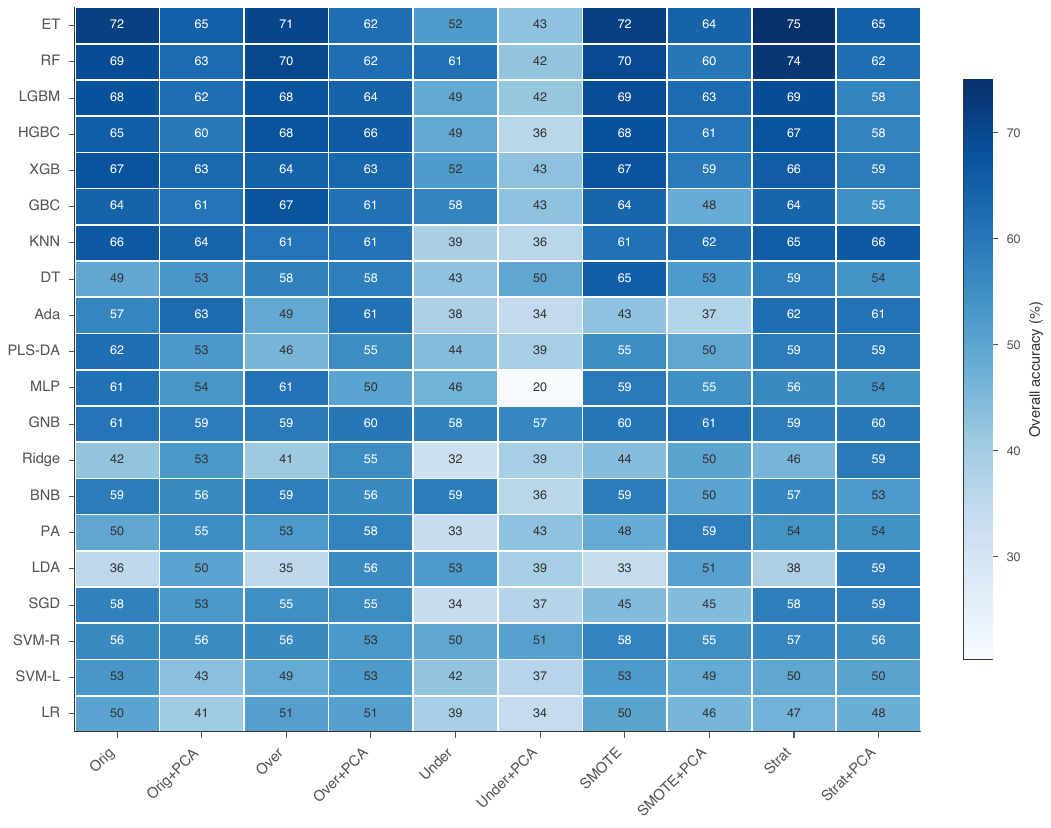}
\caption{Phase~2 preprocessing ablation across 200 configurations. Overall accuracy (\%) for 20 models (rows, sorted by best OA) across 10 preprocessing configurations (5 balancing strategies $\times$ 2 PCA settings). Vertical white lines separate balancing strategies; within each pair, the left column is without PCA and the right is with PCA. The stratified resplit / no PCA column consistently achieves the highest values.}
\label{fig:ablation}
\end{figure*}

PCA degraded performance in four of five balancing conditions. Across all 200 configurations, models without PCA outperformed those with PCA by 2.46 percentage points on average (55.62\% vs 53.16\%, Fig.~\ref{fig:ablation}). Only 7 of 20 models utilized PCA in their best Phase~2 configuration, and these were predominantly linear models (Ridge, LDA, SGDClassifier) rather than tree-based methods. The full 200-configuration landscape is visualized in Fig.~\ref{fig:ablation}, where the stratified resplit / no PCA column consistently achieves the warmest values.

\begin{table}[htbp!]
\centering
\caption{Phase~2 ablation: mean $\pm$ SD overall accuracy (\%)$\uparrow$ across 20 models by balancing strategy and PCA setting. $\Delta$ = no-PCA minus PCA difference. Best strategy in bold. Large SDs (4--11~pp) reflect substantial model-to-model variance within each configuration. $^\dagger$Stratified resplit replaces the training set ($n$=414 vs $n$=381); strategies below the rule augment the original training set.}
\label{tab:ablation}
\small
\begin{tabular}{@{}lrrr@{}}
\toprule
Balancing strategy & No PCA & PCA & $\Delta$ (pp) \\
\midrule
\textbf{Stratified resplit}$^\dagger$ & \textbf{59.08$\pm$9.26} & \textbf{57.41$\pm$4.57} & +1.67 \\
\midrule
SMOTE & 57.14$\pm$10.61 & 53.84$\pm$7.18 & +3.30 \\
Original unbalanced & 58.32$\pm$9.38 & 56.29$\pm$6.59 & +2.03 \\
Random oversampling & 57.01$\pm$9.85 & 58.12$\pm$4.47 & $-$1.11 \\
Random undersampling & 46.56$\pm$8.93 & 40.13$\pm$7.47 & +6.43 \\
\midrule
\textit{Overall mean} & 55.62 & 53.16 & +2.46 \\
\bottomrule
\end{tabular}
\end{table}

The consistent PCA degradation is surprising given its routine use in hyperspectral analysis and warrants mechanistic explanation.

Across the 20$\times$10 Phase~2 accuracy matrix, the median accuracy range attributable to preprocessing (varying balancing strategy and PCA within each model) is 25.2~pp, compared with 32.4~pp for algorithm selection (varying the model within each preprocessing configuration). A bootstrap analysis (10{,}000 resamples) yields a 95\% CI of [$-$5.8, 21.9]~pp on the difference between these median ranges, an interval that contains zero. Preprocessing and algorithm choice are thus of comparable magnitude on this benchmark. Ridge exemplifies the upper bound of preprocessing impact: its overall accuracy rose from 42.4\% (Phase~1 baseline) to 59.4\% (Phase~2 best), a 17.0~pp gain from data engineering alone.

\FloatBarrier
\subsection{Task Asymmetry and Modality Comparison}

\begin{figure*}[htbp!]
\centering
\includegraphics[width=\textwidth]{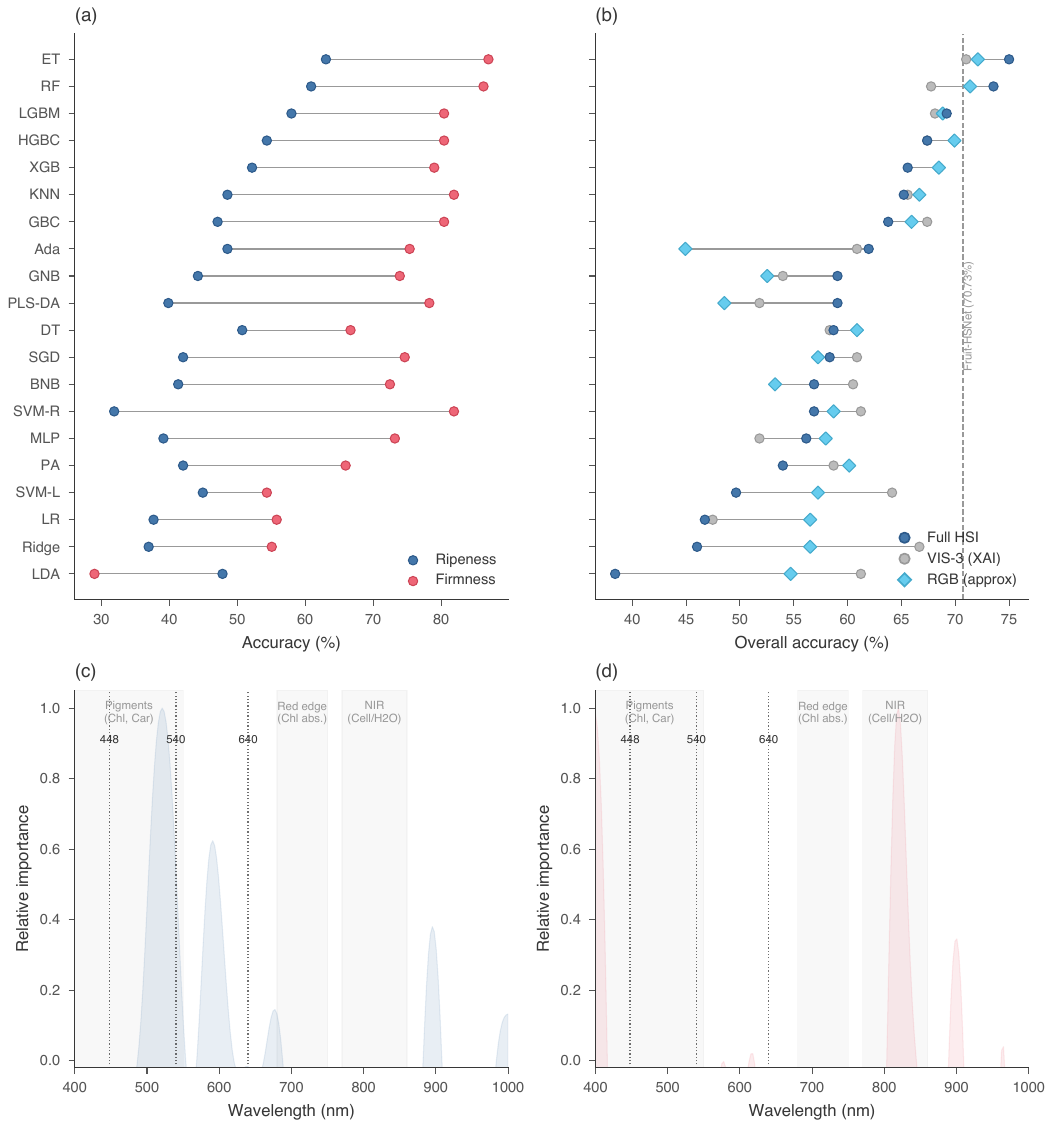}
\caption{Task asymmetry, modality comparison, and spectral importance for Phase~3 optimized models. \textbf{(a)} Task asymmetry: ripeness (pink) vs firmness (green) accuracy per model, revealing a persistent gap across all 20 algorithms. \textbf{(b)} Modality comparison: full-spectrum HSI (blue) vs VIS-3 subset (orange) overall accuracy, with the Fruit-HSNet benchmark (70.73\%) shown as a dashed line. \textbf{(c,\,d)} Spectral band importance (rolling mean $\pm$ std) for ExtraTrees across 400--1000\,nm; shaded regions highlight biologically meaningful zones (pigment changes, red-edge chlorophyll absorption, NIR cellular/water signatures). Ripeness importance concentrates in visible wavelengths (c), while firmness importance shifts toward NIR features (d).}
\label{fig:asymmetry}
\end{figure*}

Table~\ref{tab:task-asymmetry} and Fig.~\ref{fig:asymmetry} reveal two consistent patterns across all 20 models: (1) firmness prediction outperforms ripeness by a wide margin (Fig.~\ref{fig:asymmetry}a), and (2) full-spectrum data outperforms VIS-3 and RGB subsets for most competitive models, but the advantage is small (Fig.~\ref{fig:asymmetry}b).

Firmness accuracy exceeded ripeness accuracy in 19 of 20 models (mean gap 25.0 percentage points; IQR: 21.4--32.8~pp; Cohen's $d = 1.84$, large effect). This asymmetry was most extreme for SVC\_RBF (31.9\% ripeness vs 81.9\% firmness, a 50.0~pp gap) and smallest for SVC\_Linear (44.9\% vs 54.3\%, a 9.4~pp gap). LDA was the only model where ripeness accuracy exceeded firmness, likely reflecting model-specific failure modes. Because the 20 models share a common training set and test set, formal significance tests for this comparison are not appropriate (Table~\ref{tab:stat-tests}).

Phase~4 fold-level F1 comparisons between HSI and RGB revealed a large full-spectrum advantage (Cohen's $d = 1.17$; Table~\ref{tab:stat-tests}), concentrated among the top-performing models. In Phase~3, VIS-3 and RGB yield comparable performance (mean OA 61.7\% vs 60.1\%; Cohen's $d = 0.27$, $p = 0.30$), with neither modality significantly outperforming the other (Table~\ref{tab:stat-tests}). Both reduced-band configurations outperformed full-spectrum on average across all 20 models (61.7\% and 60.1\% vs 59.1\%), but this average is driven by weak models (Ridge, LDA, and SVC\_Linear) that perform substantially better in the low-dimensional 15-feature space, likely because dimensionality reduction mitigates the multicollinearity that destabilizes their coefficient estimates. At the top of the leaderboard, the full-spectrum advantage persists: ExtraTrees achieves 75.0\% (full) vs 72.1\% (RGB) vs 71.0\% (VIS-3). The persistence of the firmness--ripeness gap across all 20 models and all three modalities suggests a property of the labels rather than of the algorithms.

\begin{table*}[htbp!]
\centering
\caption{Task asymmetry and modality comparison of Phase~3 models. RA = ripeness accuracy, FA = firmness accuracy, OA = overall accuracy. Full-spectrum has 1{,}120 features; VIS-3 and RGB each use 15 features. $\Delta$OA = Full OA $-$ max(VIS-3, RGB). %\textbf{Bold} = best in column; \ul{underline} = second-best. Arrows indicate preferred direction.
}
\label{tab:task-asymmetry}
\small
\begin{tabular}{@{}l l rrr rr r@{}}
\toprule
& & \multicolumn{3}{c}{Full-spectrum} & \multicolumn{2}{c}{Reduced-band OA$\uparrow$} & \\
\cmidrule(lr){3-5} \cmidrule(lr){6-7}
Rk & Model & RA$\uparrow$ & FA$\uparrow$ & OA$\uparrow$ & VIS-3 & RGB & $\Delta$OA \\
\midrule
1 & ExtraTrees & \textbf{63.0} & \textbf{87.0} & \textbf{75.0} & \textbf{71.0} & \textbf{72.1} & +2.9 \\
2 & RandomForest & \ul{60.9} & \ul{86.2} & \ul{73.6} & 67.8 & \ul{71.4} & +2.2 \\
3 & LGBM & 58.0 & 80.4 & 69.2 & 68.1 & 68.8 & +0.4 \\
4 & HistGradBoosting & 54.3 & 80.4 & 67.4 & 67.4 & 69.9 & $-$2.5 \\
5 & XGBoost & 52.2 & 79.0 & 65.6 & \ul{68.5} & 68.5 & $-$2.9 \\
6 & KNeighbors & 48.6 & 81.9 & 65.2 & 65.6 & 66.7 & $-$1.4 \\
7 & GradientBoosting & 47.1 & 80.4 & 63.8 & 67.4 & 65.9 & $-$3.6 \\
8 & AdaBoost & 48.6 & 75.4 & 62.0 & 60.9 & 44.9 & +1.1 \\
9 & GaussianNB & 44.2 & 73.9 & 59.1 & 54.0 & 52.5 & +5.1 \\
10 & PLSDA & 39.9 & 78.3 & 59.1 & 51.8 & 48.6 & +7.2 \\
11 & DecisionTree & 50.7 & 66.7 & 58.7 & 58.3 & 60.9 & $-$2.2 \\
12 & SGDClassifier & 42.0 & 74.6 & 58.3 & 60.9 & 57.2 & $-$2.5 \\
13 & BernoulliNB & 41.3 & 72.5 & 56.9 & 60.5 & 53.3 & $-$3.6 \\
14 & SVC\_RBF & 31.9 & 81.9 & 56.9 & 61.2 & 58.7 & $-$4.3 \\
15 & MLPClassifier & 39.1 & 73.2 & 56.2 & 51.8 & 58.0 & $-$1.8 \\
16 & PassiveAggr. & 42.0 & 65.9 & 54.0 & 58.7 & 60.1 & $-$6.2 \\
17 & SVC\_Linear & 44.9 & 54.3 & 49.6 & 64.1 & 57.2 & $-$14.5 \\
18 & LogisticRegr. & 37.7 & 55.8 & 46.7 & 47.5 & 56.5 & $-$9.8 \\
19 & Ridge & 37.0 & 55.1 & 46.0 & 66.7 & 56.5 & $-$20.7 \\
20 & LDA & 47.8 & 29.0 & 38.4 & 61.2 & 54.7 & $-$22.8 \\
\midrule
\textit{Mean} & & 46.6 & 71.6 & 59.1 & 61.7 & 60.1 & -4.0 \\
\bottomrule
\end{tabular}
\end{table*}

\FloatBarrier
\subsection{Per-Fruit, Per-Class, and Ensemble Analysis}

Table~\ref{tab:per-fruit} reports per-fruit accuracy for the top five models (Fig.~\ref{fig:per_fruit} in Appendix), revealing substantial variation across species. Avocado and mango achieved highest accuracies (up to 75.0\% for mango with RandomForest), while papaya consistently underperformed (33.3--44.4\%), likely due to its smallest sample size ($n$=51).

\begin{table*}[htbp!]
\centering
\caption{Per-fruit ripeness accuracy (\%) for the top-five models (Phase~4 single-split holdout, $n$=138; ranked by Phase~3 overall accuracy). \textbf{Bold} = best per fruit; \ul{underline} = second-best.}
\label{tab:per-fruit}
\small
\begin{tabular}{@{}lrrrrr@{}}
\toprule
Fruit ($n$) & ExtraTrees & RandomForest & LightGBM & HistGradBoosting & XGBoost \\
\midrule
Avocado (170) & \textbf{67.9} & \ul{65.4} & 44.4 & 48.1 & 38.3 \\
Kaki (68) & \textbf{41.7} & \textbf{41.7} & 33.3 & \textbf{41.7} & \textbf{41.7} \\
Kiwi (162) & 58.3 & 54.2 & \ul{62.5} & \textbf{70.8} & 50.0 \\
Mango (68) & \ul{66.7} & \textbf{75.0} & 41.7 & \ul{66.7} & 50.0 \\
Papaya (51) & \ul{33.3} & \ul{33.3} & \ul{33.3} & \ul{33.3} & \textbf{44.4} \\
\bottomrule
\end{tabular}
\end{table*}

\begin{figure*}[htbp!]
\centering
\includegraphics[width=\textwidth]{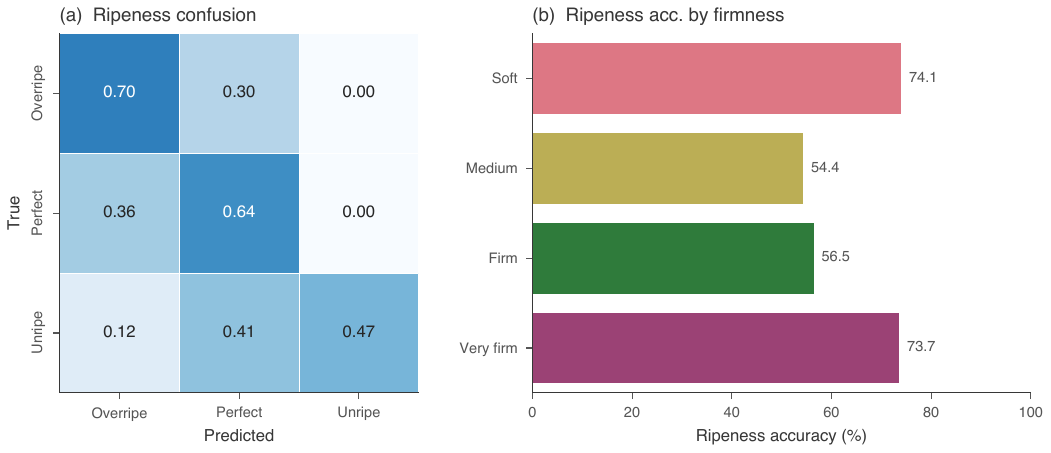}
\caption{Classification patterns for the best model (ExtraTrees, single-split holdout, $n$=138). \textbf{(a)} Normalized ripeness confusion matrix showing primary confusion between adjacent maturation states. \textbf{(b)} Ripeness accuracy stratified by firmness group (4-bin); extreme-firmness groups achieve higher ripeness accuracy than intermediate groups.}
\label{fig:confusion}
\end{figure*}

Table~\ref{tab:per-class} provides per-class precision, recall, and F1-score from the single-split holdout evaluation. Ripeness classification showed primary confusion between adjacent states (unripe/perfect and perfect/overripe), as visible in the normalized confusion matrix (Fig.~\ref{fig:confusion}a). Stratifying ripeness accuracy by firmness group reveals that extreme-firmness samples (soft: 74.1\%, very firm: 73.7\%) are classified more accurately than intermediate groups (medium: 54.4\%, firm: 56.5\%; Table~\ref{tab:per-class}, Fig.~\ref{fig:confusion}b), suggesting that fruits at firmness extremes present clearer ripeness signatures.

Ten-fold stratified cross-validation (Fig.~\ref{fig:cv_stability}, Fig.~\ref{fig:fold-dist} in Appendix) confirmed that top models generalize reliably. ExtraTrees achieved $82.2 \pm 4.7\%$ mean OA, with narrow 95\% confidence intervals. Firmness accuracy was consistently higher than ripeness across all models, with tighter variance. This pattern reinforces the task asymmetry observed in Phase~3.

\begin{table}[htbp!]
\centering
\caption{ExtraTrees Phase~4 single-split holdout analysis ($n$=138). \textbf{Top}: ripeness per-class metrics from the confusion matrix (Fig.~\ref{fig:confusion}a). \textbf{Bottom}: ripeness accuracy stratified by firmness group (4-bin analysis on the original unbalanced split). Extreme-firmness groups achieve higher ripeness accuracy than intermediate groups. $^*$Unripe precision~(1.000) with 47.1\% recall means the model rarely predicts unripe; most unripe fruit are misclassified as perfect (41.2\%) or overripe (11.8\%).}
\label{tab:per-class}
\small
\begin{tabular}{@{}llrrr@{}}
\toprule
Task & Class & Precision & Recall & F1 \\
\midrule
Ripeness & Overripe & 0.561 & \textbf{0.696} & 0.621 \\
Ripeness & Perfect & \ul{0.569} & \ul{0.638} & \ul{0.602} \\
Ripeness & Unripe$^*$ & \textbf{1.000} & 0.471 & \textbf{0.640} \\
\midrule
\multicolumn{2}{@{}l}{\textit{Firmness group}} & \multicolumn{3}{c}{\textit{Ripeness accuracy}} \\
\cmidrule(l){3-5}
& Soft ($n$=27) & \multicolumn{3}{c}{0.741} \\
& Medium ($n$=68) & \multicolumn{3}{c}{0.544} \\
& Firm ($n$=23) & \multicolumn{3}{c}{0.565} \\
& Very firm ($n$=19) & \multicolumn{3}{c}{\textbf{0.737}} \\
\bottomrule
\end{tabular}
\end{table}

\begin{figure*}[htbp!]
\centering
\includegraphics[width=\textwidth]{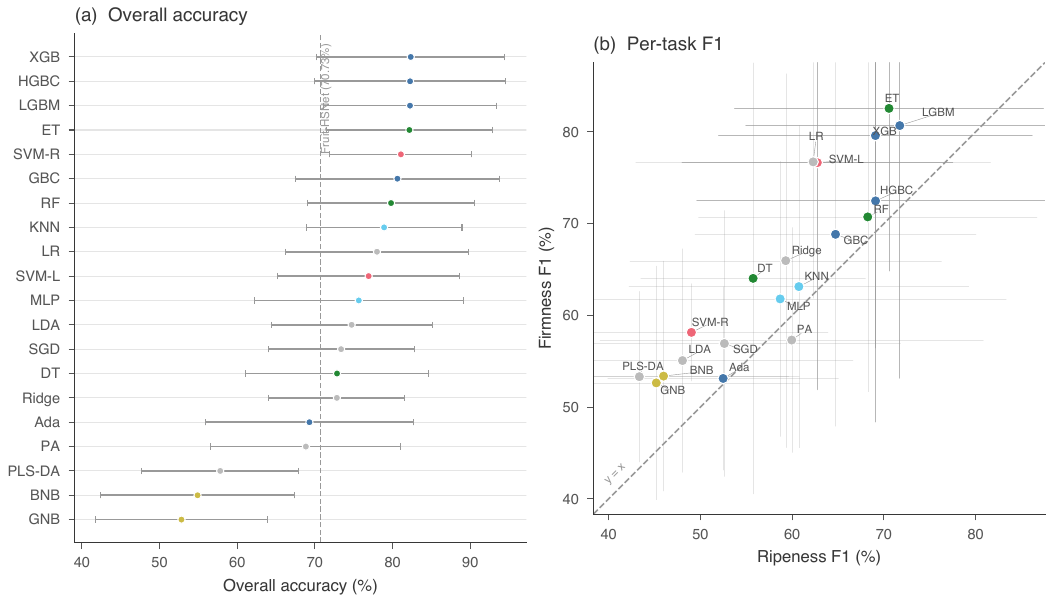}
\caption{Cross-validation stability (Phase~4, 10-fold stratified CV). \textbf{(a)}~Overall accuracy with 95\% confidence intervals (mean $\pm$ $t_{9,0.025}$$\cdot$std, where $t_{9,0.025} \approx 2.262$) for all 20 models, sorted by mean. \textbf{(b)}~Task-separated view: ripeness vs firmness accuracy with CIs. Firmness predictions are consistently more accurate and stable.}
\label{fig:cv_stability}
\end{figure*}

\begin{figure}[htbp!]
\centering
\includegraphics[width=\columnwidth]{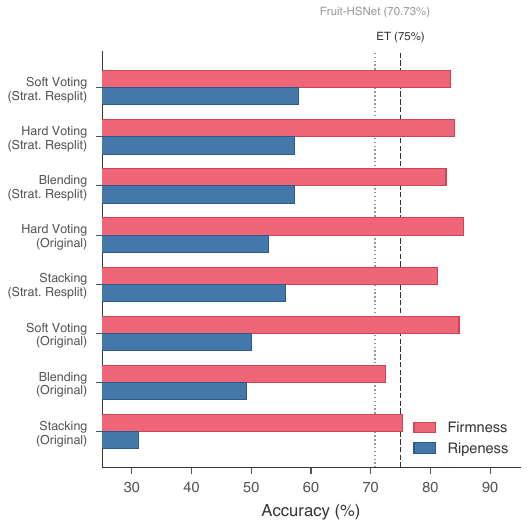}
\caption{Ensemble performance decomposition by task. All eight ensemble configurations maintain high firmness accuracy ($>$72\%) but catastrophically degrade ripeness accuracy (stacking drops to 31.2\%). Dashed line: ExtraTrees single-model baseline (75.0\%); dotted line: Fruit-HSNet benchmark (70.73\%).}
\label{fig:ensemble}
\end{figure}

\FloatBarrier
Ensembles combined the top five base learners using four aggregation strategies under original unbalanced and optimal preprocessing configurations but did not surpass the best single model in either modality (Table~\ref{tab:ensemble}, Fig.~\ref{fig:ensemble}). The best full-spectrum ensemble (hard/soft voting with stratified resplit) achieved 70.65\% overall accuracy versus 75.00\% for ExtraTrees alone, a 4.35 percentage point degradation.

\begin{table}[htbp!]
\centering
\caption{Ensemble performance comparison. All ensembles use the top-5 models from Phase~3. $\Delta$ET shows difference from best single model (ExtraTrees, 75.0\%).}
\label{tab:ensemble}
\small
\setlength{\tabcolsep}{4pt}
\begin{tabular}{llcccc}
\toprule
Strategy & Data State & Ripe$\uparrow$ & Firm$\uparrow$ & OA$\uparrow$ & $\Delta$ET \\
\midrule
  Hard Voting & Original & 52.9 & 85.5 & 69.2 & -5.8 \\
  Soft Voting & Original & 50.0 & 84.8 & 67.4 & -7.6 \\
  Stacking & Original & 31.2 & 75.4 & 53.3 & -21.7 \\
  Blending & Original & 49.3 & 72.5 & 60.9 & -14.1 \\
\midrule
  Hard Voting & Strat. Resplit & 57.2 & 84.1 & 70.7 & -4.3 \\
  Soft Voting & Strat. Resplit & 58.0 & 83.3 & 70.7 & -4.3 \\
  Stacking & Strat. Resplit & 55.8 & 81.2 & 68.5 & -6.5 \\
  Blending & Strat. Resplit & 57.2 & 82.6 & 69.9 & -5.1 \\
\bottomrule
\end{tabular}
\end{table}

\FloatBarrier
\subsection{Explainable AI Analysis}

\begin{figure*}[htbp!]
\centering
\includegraphics[width=\textwidth]{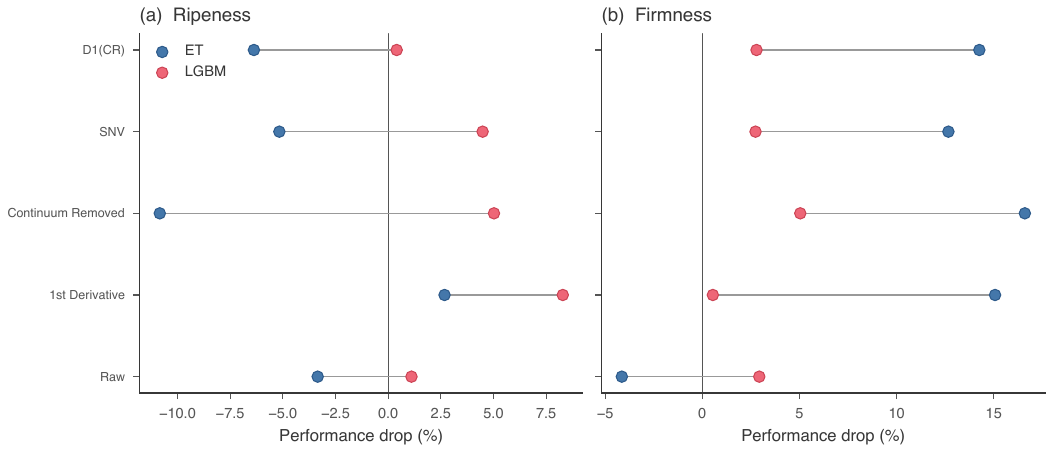}
\caption{Feature group ablation: relative performance drop (\%) when each spectral transformation group (224 bands) is removed. \textbf{(a)} Ripeness: first derivatives are most critical for LightGBM (8.3\% drop); other groups are redundant or harmful for ExtraTrees. \textbf{(b)} Firmness: continuum-removed features dominate ExtraTrees (16.6\% drop), while LightGBM shows uniform sensitivity across groups ($\leq$5.0\%). Positive values indicate removing the group hurts performance; negative values indicate removal improves it. Task- and model-dependent contributions explain why PCA, which compresses all groups uniformly, degrades classification.}
\label{fig:feature_groups}
\end{figure*}

Phase~6 XAI analysis for ExtraTrees revealed concentrated spectral band importance in three regions (Fig.~\ref{fig:asymmetry}c,\,d): visible blue/green (400--550~nm) associated with pigment changes, red-edge transitions ($\sim$700--750~nm) indicating vegetation stress, and near-infrared ($\sim$770--860~nm) correlated with cellular structure. Feature group ablation (Fig.~\ref{fig:feature_groups}) confirmed that the five spectral transformations contribute differently across tasks and models. For ripeness, first derivatives are the most critical group for LightGBM (8.3\% drop when removed), while other preprocessing groups are redundant or harmful for ExtraTrees (removing continuum-removed features improves performance by 10.8\%). For firmness, continuum-removed features dominate ExtraTrees (16.6\% drop), while all five groups contribute more uniformly to LightGBM (largest drop 5.0\%). These task- and model-dependent feature contributions explain why the full 1{,}120-dimensional feature space outperforms PCA reduction: each transformation captures complementary discriminative information that varies by task.

Ripeness prediction drew primarily on visible-spectrum features associated with chlorophyll degradation and carotenoid accumulation, while firmness prediction relied more heavily on near-infrared features correlated with cellular structure and water content. This wavelength-task correspondence aligns with established plant physiology knowledge \citep{khodabakhshian2017application, feng2023nondestructive}. However, XAI attributions may partly reflect species-level spectral differences rather than within-species quality-state variation. The PCA and t-SNE projections (Fig.~\ref{fig:dataset}c,d) show that fruit type dominates the variance structure, so wavelength regions that separate species could receive high importance scores even if they carry limited quality information within a single species.

\section{Discussion}
\label{sec:discussion}

The comprehensive six-phase evaluation reveals three recurring patterns across 20 algorithms applied to ripeness classification and firmness prediction across five fruit species. Nevertheless, the results raise three fundamental questions that span all evaluated approaches. What mechanism makes preprocessing as influential as algorithm choice? Why does the firmness-ripeness gap persist regardless of model family or modality? And why does PCA, a standard step in hyperspectral workflows, consistently degrade accuracy on this feature space?
We discuss each in turn before addressing methodological observations and limitations.

\subsection{The Interplay of Preprocessing and Algorithm Choice}

The relative importance of preprocessing versus algorithm choice depends on the model family. A bootstrap analysis (10{,}000 resamples) of the 20$\times$10 Phase~2 accuracy matrix yields a 95\% CI of [$-$5.8, 21.9]~pp on the difference between the two median ranges (algorithm: 32.4~pp; preprocessing: 25.2~pp), an interval that contains zero. Neither source of variance can be declared the unambiguous dominant factor with confidence.

\begin{figure*}[htbp!]
\centering
\includegraphics[width=\textwidth]{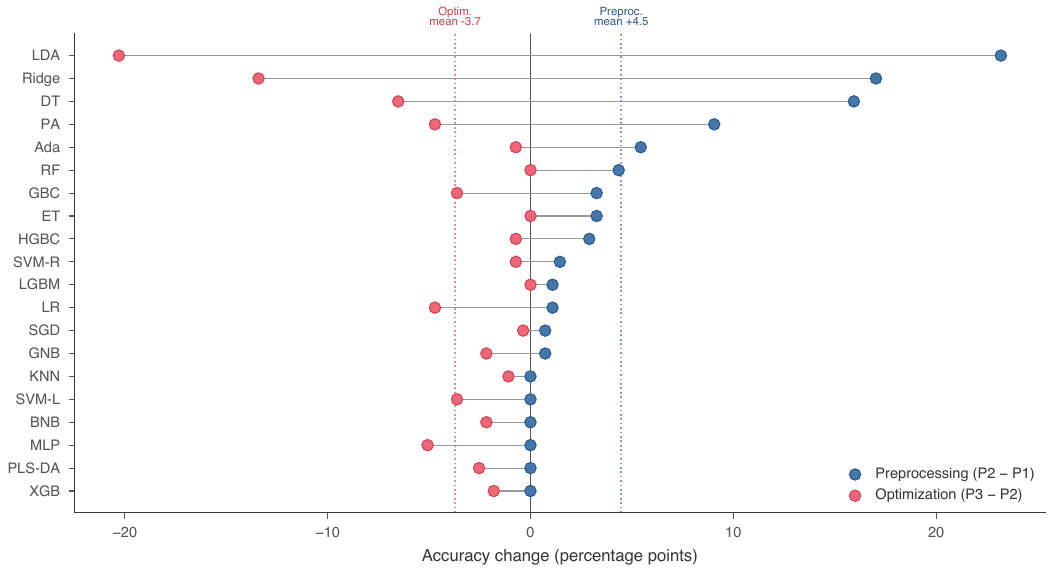}
\caption{Preprocessing gain vs optimization gain for all 20 models. Each model's accuracy improvement is decomposed into preprocessing gain (best Phase~2 config minus Phase~1 baseline, teal) and optimization gain (Phase~3 minus best Phase~2, purple). Models sorted by preprocessing gain. Dotted lines indicate means. For most models, preprocessing contributes the dominant share of improvement; optimization gain is near-zero or negative for many.}
\label{fig:preprocess-dominance}
\end{figure*}

For the best-performing models, however, preprocessing is the dominant lever: the best preprocessing configuration improved 15 of 20 models over their Phase~1 baselines, while Phase~3 optimization did not improve any model beyond its best Phase~2 result. For weaker models (LDA, Ridge), preprocessing gains are larger in absolute terms (up to 23~pp) but algorithm choice matters more; these models remain far below competitive accuracy regardless of preprocessing. Fig.~\ref{fig:preprocess-dominance} decomposes each model's accuracy trajectory into preprocessing gain (Phase~1$\to$Phase~2) and optimization gain (Phase~2$\to$Phase~3), directly visualizing this asymmetry.

The mechanism behind stratified resplit's advantage over synthetic resampling (SMOTE, random oversampling) likely stems from the structure of spectral data. SMOTE generates new samples by interpolating between existing ones in 1{,}120-dimensional feature space, an operation that assumes local linearity in spectral manifolds shaped by nonlinear biochemical processes. Interpolating between two spectra at different maturation stages does not necessarily produce a spectrum corresponding to an intermediate state. Stratified resplit avoids this problem by redistributing genuine samples across train and test splits without fabrication. However, the stratified resplit condition also increases training set size by 8.7\% (414 vs 381 samples) and improves species balance, so its advantage may partly reflect additional data and improved species representation rather than purely a balancing effect.

The practical implication is that for competitive models, preprocessing investment yields larger returns than hyperparameter tuning, while algorithm selection establishes the performance ceiling. Most published studies on this dataset evaluate a single preprocessing pipeline without systematic ablation. Our Phase~2 results across 200 configurations (Fig.~\ref{fig:ablation}) demonstrate that varying the preprocessing pipeline alone shifts accuracy by up to 23~pp for the same algorithm (LDA: 35.5\% to 58.7\%), which may obscure the true source of performance variation in published comparisons.

\subsection{Task Asymmetry as a Construct Validity Question}

The persistent firmness--ripeness gap is not simply a matter of task difficulty; it reflects a difference in what the two labels measure.

Firmness is a continuous mechanical property with direct spectral correlates. Within the limited NIR range of the FX10 sensor (770--960~nm), absorption features respond to changes in cellular structure and water content associated with tissue softening \citep{lu2020recent, khodabakhshian2017application}. Discretizing firmness into three classes (soft, medium, firm) creates boundaries that align with real physical thresholds. Ripeness, by contrast, is a composite construct. A single ripeness label aggregates chlorophyll degradation, carotenoid accumulation, sugar-acid balance, volatile production, and textural changes, processes that occur at different rates and may produce conflicting spectral signatures within the same fruit. The confusion between adjacent ripeness states (overripe/perfect: 33.3\% mean bidirectional misclassification, unripe$\to$perfect: 41.2\%; Fig.~\ref{fig:confusion}a) reflects the inherent ambiguity of discretizing a multi-dimensional process into ordinal categories.

The persistence of this gap across all 20 models and both modalities suggests the asymmetry is a property of the labeling scheme rather than of specific algorithms or sensors. \citet{feng2023nondestructive} predicted multiple loquat quality parameters and found that prediction accuracy varied across attributes (color $R^2_P = 0.96$, firmness $R^2_P = 0.87$, SSC $R^2_P = 0.84$), consistent with the view that different quality dimensions have inherently different spectral predictability. Future work should investigate whether continuous regression targets or multi-dimensional maturity indices reduce the information loss inherent in categorical ripeness labels.

For deployment, the asymmetry is more acceptable than it first appears. Firmness directly determines shelf life and consumer satisfaction, and commercial distributors prioritize rejecting unacceptably soft fruit over classifying exact ripeness state. However, per-species firmness accuracy (Table~\ref{tab:per-fruit}) remains below 70\% for most individual fruit types at Phase~4, which is insufficient for fully autonomous commercial sorting without human oversight. The results establish a useful benchmark but should not be interpreted as a deployable system specification.

\subsection{PCA Degradation and the Variance--Discrimination Mismatch}

The consistent PCA degradation contradicts standard practice in hyperspectral analysis \citep{lu2020recent, ahmad2022comprehensive}. The explanation is rooted in our feature engineering pipeline.

The 1{,}120-feature space concatenates five spectral transformations, each designed to isolate different physical phenomena: raw reflectance captures absolute intensity, first derivatives capture absorption edge slopes, continuum removal normalizes band depth, and SNV corrects for scattering. PCA maximizes variance across the concatenated space, but variance and task-relevant discrimination are not aligned. First derivatives contribute $<$0.1\% of total variance across the 1{,}120 features, yet removing them drops LightGBM ripeness accuracy by 8.3\% (Fig.~\ref{fig:feature_groups}a). PCA at 95\% retained variance effectively discards these low-variance, high-discrimination features.

The 7 models that selected PCA in their optimal Phase~2 configuration were predominantly linear (Ridge, LDA, SGDClassifier, PassiveAggressive). These methods benefit from PCA because dimensionality reduction mitigates the multicollinearity that destabilizes their coefficient estimates. Tree-based methods face no such constraint: their split-based decision boundaries are invariant to feature correlation. This interaction between model family and preprocessing underlines why blanket preprocessing recommendations of always applying PCA to hyperspectral data can be counterproductive when engineered features precede dimensionality reduction.

\subsection{Full-Spectrum Versus Reduced-Band Sensing}

The full-spectrum advantage at the top of the leaderboard is modest: ExtraTrees achieves 75.0\% with 1{,}120 features compared to 72.1\% with approximate RGB bands (96.1\% recovery) and 71.0\% with XAI-derived VIS-3 bands (94.7\% recovery). Hyperspectral systems cost \$10{,}000--100{,}000 compared to \$500--5{,}000 for multispectral or RGB alternatives \citep{min2023decay}. Both reduced-band configurations recover 94--96\% of full-spectrum accuracy using 15 features, and several models achieve comparable or better Phase~3 OA in the low-dimensional space (Table~\ref{tab:task-asymmetry}).

The VIS-3 and RGB configurations produce statistically indistinguishable results across 20 models (mean OA 61.7\% vs 60.1\%; Cohen's $d = 0.27$, $p = 0.30$; Table~\ref{tab:stat-tests}). The two band sets differ by only 1--4 spectral indices (2--11\,nm in wavelength), yet neither configuration holds a consistent advantage. This insensitivity to exact band placement indicates that the discriminative spectral content in the visible range is spectrally broad: any three well-separated bands within the 400--700\,nm window capture it. XAI-guided band selection provides no measurable benefit over generic RGB-center wavelengths on this benchmark.

XAI analysis concentrates discriminative wavelengths in three regions: visible blue/green (400--550~nm) for pigment assessment, the red edge ($\sim$700--750~nm) for chlorophyll transitions, and near-infrared ($\sim$770--860~nm) for cellular structure (Fig.~\ref{fig:asymmetry}c,d). This concentration aligns with the spectral-biochemical relationships established by \citet{khodabakhshian2017application} and \citet{feng2023nondestructive}. However, interpretations specific to the short-NIR boundary (770--960~nm) should be treated as speculative relative to the well-established core NIR firmness bands (1200--1450~nm), which fall outside the Specim FX10 range; the firmness-related importance observed here may partly capture species-level variance rather than within-species quality differences.

Two caveats apply. First, both reduced-band configurations extract features from hyperspectral cubes at exact wavelengths, bypassing the spectral response curves, Bayer filter artifacts, and noise characteristics of real sensors. The reduced-band accuracy figures represent upper bounds. That said, the RGB parity result weakens the ``upper bound'' framing: if approximate RGB wavelengths match XAI-optimized bands, the gap between simulated and real sensor performance may be smaller than the VIS-3 result alone suggests. Second, the full-spectrum advantage may increase for tasks requiring finer spectral discrimination, such as early-stage decay detection where subtle NIR absorption shifts precede visible symptoms \citep{min2023decay}.

\subsection{Tree-Based Dominance and Ensemble Underperformance}

Tree-based ensembles (ExtraTrees, RandomForest) occupied the top two positions in Phase~3 and remained competitive in Phase~4 cross-validation ($82.2 \pm 4.7\%$ and $79.8 \pm 4.8\%$). Their advantage arises from a structural match with the feature space. The 1{,}120 features comprise five transformation groups (224 bands each) with high within-group correlation but distinct cross-group discriminative content (Fig.~\ref{fig:feature_groups}). ExtraTrees' random split-point selection \citep{geurts2006extremely} acts as implicit random subspace sampling across these groups, decorrelating individual trees without explicit feature selection. Linear methods face a different problem: multicollinearity among the 224 correlated bands within each transformation inflates coefficient variance and degrades generalization.

The ensemble underperformance with homogeneous base learners (Table~\ref{tab:ensemble}) is instructive. All four aggregation strategies degraded performance relative to ExtraTrees alone, with stacking suffering the worst drop ($-21.7$~pp under original data). The top-five base learners (ExtraTrees, RandomForest, LGBM, HistGradientBoosting, XGBoost) are all recursive partitioning methods that construct decision boundaries through axis-aligned splits. They make correlated errors, particularly on the same borderline ripeness cases (Table~\ref{tab:per-class}). Ensemble theory predicts that aggregation improves predictions only when base learner errors are sufficiently uncorrelated. Including a structurally different learner (e.g., SVM or a neural network) as a base learner might restore the diversity that ensemble methods require, though at the cost of interpretability.

ExtraTrees, despite providing feature importance scores through tree-based splitting, is not fully transparent: predictions for individual samples cannot be explained without post-hoc XAI tools. The same limitation applies to all gradient-boosted methods evaluated here. The interpretability advantage of tree-based methods is relative to black-box neural networks, not absolute.

\subsection{Spatial Complexity and Mean-Spectrum Representations}

ExtraTrees achieves 75.0\% on the benchmark split using only per-sample mean spectra, a representation that retains spectral wavelength information but discards all spatial variation across fruit pixels. Fruit-HSNet's dual-branch architecture (Section~\ref{sec:methods}) processes full spatial cubes yet reports 70.73\% under a different protocol. Two interpretations are consistent with this outcome. Spatial variation across fruit pixels may carry little additional discriminative signal beyond what the mean spectrum encodes. Alternatively, mean spectrum aggregation may lose useful spatial information, but any advantage Fruit-HSNet derives from recovering it may be offset by other architectural differences. Distinguishing these interpretations requires ablation within the Fruit-HSNet framework, which the available data cannot support. The practical implication holds regardless of cause. ExtraTrees trains in 0.2{} seconds on a consumer CPU, requires no data augmentation, produces a 2.77~MB model, and achieves 2.1~ms per-sample inference latency.

The three discriminative wavelengths identified by explainability analysis (448, 540, 640\,nm)
were selected as the top joint-ranked visible-range bands from ExtraTrees consensus feature
importance across both tasks (Fig.~\ref{fig:asymmetry}c,d, dotted lines). These wavelengths
lie within 2--11\,nm of standard RGB filter centers (450, 550, 650\,nm), and the controlled
comparison confirms that this proximity translates to equivalent classification performance:
RGB-center bands recover 96.1\% of full-spectrum accuracy versus 94.7\% for VIS-3
(Table~\ref{tab:task-asymmetry}). Commodity RGB cameras use broad, overlapping passbands
rather than narrow spectral selections, but the insensitivity to exact band placement observed
here suggests that such cameras may capture sufficient discriminative content for this
task. Validation with actual RGB sensor data remains necessary to confirm this implication,
because Bayer-pattern demosaicing, sensor noise, and spectral crosstalk introduce artifacts
absent from simulated band selection.

The spectroscopic basis of these feature importance findings carries direct implications for purpose-built inline sensing hardware. Discriminative information concentrated near the chlorophyll absorption trough ($\sim$680\,nm), the red-edge transition (700--750\,nm), and the tissue water/structure region (770--860\,nm) suggests that a targeted multispectral filter array covering these regions could capture a substantial portion of the discriminative spectral content available in the full 224-band dataset, at a fraction of hyperspectral system cost. Reducing to 6--12 strategically positioned channels would lower per-sample data volume sufficiently to enable inference on embedded controllers without GPU acceleration, a key deployment constraint in high-speed sorting environments. The demonstrated importance of SNV normalization, illustrated by the 12.7\% relative performance drop in ExtraTrees firmness upon its removal (Fig.~\ref{fig:feature_groups}b), is not merely a preprocessing detail but a spectroscopically motivated robustness measure against the multiplicative scatter effects of variable surface geometry and path length inherent to production-line acquisition. On a reduced-band sensor, however, computing SNV statistics across only a handful of spectral points becomes poorly conditioned; alternative scatter correction strategies such as external diffuse reference tiles or integrating sphere-based calibration panels would be required to preserve this robustness in practice.

Push-broom line-scan sensors are architecturally well-matched to conveyor belt translation, but rotating or irregularly presented fruit introduces variable illumination geometry that complicates line-scan acquisition. Snapshot multispectral mosaic sensors, which acquire full spatial and spectral information in a single frame, may be better suited to this geometry at the cost of lower spectral resolution, and represent a complementary hardware direction worth evaluating against the wavelength importance findings reported here.

\subsection{Single-Split Versus Cross-Validated Evaluation}

A pattern in Table~\ref{tab:comprehensive} warrants attention: several models rank very differently in Phase~3 (single train-test split) versus Phase~4 (10-fold CV). SVC\_RBF achieves only 56.9\% in Phase~3 but $81.1 \pm 4.0\%$ in Phase~4. LogisticRegression drops to 46.7\% in Phase~3 while reaching $78.0 \pm 5.2\%$ in Phase~4. Conversely, GaussianNB scores 59.1\% in Phase~3 but only $52.8 \pm 4.9\%$ in Phase~4.

These discrepancies arise because Phase~3 evaluates on a single held-out set, making results sensitive to the specific composition of that split. Phase~4 cross-validation averages over 10 splits and provides confidence intervals. The models most affected (SVMs, LogisticRegression, LDA) are those whose decision boundaries are sensitive to the training sample composition. Tree-based ensembles show smaller discrepancies because bagging and random subspace sampling provide internal averaging that reduces split sensitivity. Even with identical hyperparameters and data splits, stochastic models such as ExtraTrees may vary across re-trainings due to internal randomization (e.g., random subspace selection); the $\sim$2~pp gap between ExtraTrees' Phase~3 holdout accuracy (75.0\%) and Phase~4 holdout re-evaluation (73.2\%) reflects this inherent variance rather than a methodological inconsistency.

This observation has a methodological implication for the field: studies reporting accuracy from a single train-test partition may rank algorithms differently than cross-validated evaluations. The Fruit-HSNet benchmark (70.73\%) was itself evaluated on a single split \citep{benjmaa2025fruit}, which limits the precision of any direct comparison. Our Phase~4 cross-validated estimates are more reliable for algorithm ranking, though they evaluate a different quantity (generalization across random partitions) than Phase~3 (performance on the dataset's original held-out set).

\subsection{Limitations}

Four limitations affect the generalizability of these findings most directly. First, all results derive from a single dataset (DeepHS Fruit) captured with one camera system (Specim FX10) under laboratory conditions across 47 acquisition days. Field environments introduce illumination variation, background clutter, and 15--30\% accuracy degradation \citep{min2023decay}. Whether the preprocessing-over-algorithm ordering holds under field conditions remains untested. Per-fruit accuracy (33--67\% for individual species at Phase~4) falls well below the threshold required for fully autonomous commercial sorting without human oversight; these results should be interpreted as a research benchmark, not a deployment claim.

Second, the comparison to Fruit-HSNet (70.73\%) involves a protocol mismatch: \citet{benjmaa2025fruit} used a different train-test partition and evaluation procedure. The Phase~3 result (ExtraTrees, 75.0\%; Wilson 95\% CI: [67.6\%, 81.8\%]) and Fruit-HSNet's reported value are not directly comparable. The ExtraTrees confidence interval contains the Fruit-HSNet value. The scikit-learn \texttt{MLPClassifier} benchmarked here is a shallow feedforward network, not a deep learning model. ExtraTrees' 75.0\% numerically exceeds Fruit-HSNet's 70.73\%, but the CI contains that value and protocol differences preclude a controlled comparison; no general superiority claim over deep learning architectures is warranted. The benchmark is scoped to the scikit-learn ecosystem.

Third, sample sizes are imbalanced across fruit species (papaya $n$=51 vs avocado $n$=170), and papaya consistently underperforms (33--44\% accuracy; Table~\ref{tab:per-fruit}). Per-fruit deployment would require species-specific calibration and substantially larger training sets for underrepresented varieties.

Fourth, PLS-DA, the standard chemometric baseline, is included but ranks last among the 20 models (P4 CV: $57.8 \pm 4.5\%$), likely because the effective number of latent components is clamped to the number of classes ($n=3$), severely limiting its capacity on the 1{,}120-feature space. This result should not be interpreted as evidence that PLS-DA is inherently inferior for spectral classification; with larger class counts or continuous regression targets, PLS-DA may perform competitively. %Additionally, the PLS-DA Phase~4 cross-validation exhibits a pipeline anomaly: per-fold ripeness and firmness accuracies are identical, indicating that the dual-task separation did not function correctly for this model. The reported P4 value (57.8\%) therefore reflects a single-task evaluation rather than a true dual-task average; PLS-DA's Phase~3 single-split result (59.1\%) is unaffected. Multiple testing across 200 configurations was not formally corrected; the best-configuration selection is subject to selection bias over the full Phase~2 search space, which Phase~4 cross-validation mitigates but does not eliminate.

\section{Conclusion}
\label{sec:conclusion}
In this study, we investigated various preprocessing strategies and machine learning algorithms for non-destructive estimation of fruit ripeness and firmness from hyperspectral imaging. Our evaluations showed that preprocessing choices can influence hyperspectral fruit quality prediction as strongly as algorithm selection, with preprocessing improvements sometimes exceeding gains from model optimization.
%In several cases, improvements from preprocessing exceeded gains obtained through algorithm optimization, indicating that systematic preprocessing ablation should precede extensive model search when developing hyperspectral analysis pipelines. %In particular, indiscriminate use of PCA may reduce performance when applied after engineered spectral features, emphasizing the need for model-specific preprocessing evaluation rather than default pipelines.
The consistent performance gap between firmness and ripeness prediction appears to arise from differences in label structure rather than algorithmic limitations.
Tree-based ensemble models achieved up to 82.2\% cross-validated accuracy on standard consumer CPU hardware. Additionally, explainable AI analysis identified three informative visible-range wavelengths that retained 94.7\% of full-spectrum performance using only 15 features, highlighting their potential for targeted multispectral sensor design. Future work should validate these findings across larger multi-species datasets to assess the generalizability of the preprocessing–algorithm relationship and the transferability of the identified wavelengths.

\appendix
\setcounter{table}{0}
\setcounter{figure}{0}

\section{Supplementary Figures}
\label{sec:appendix-figs}

\begin{figure*}[htbp!]
\centering
\includegraphics[width=\textwidth]{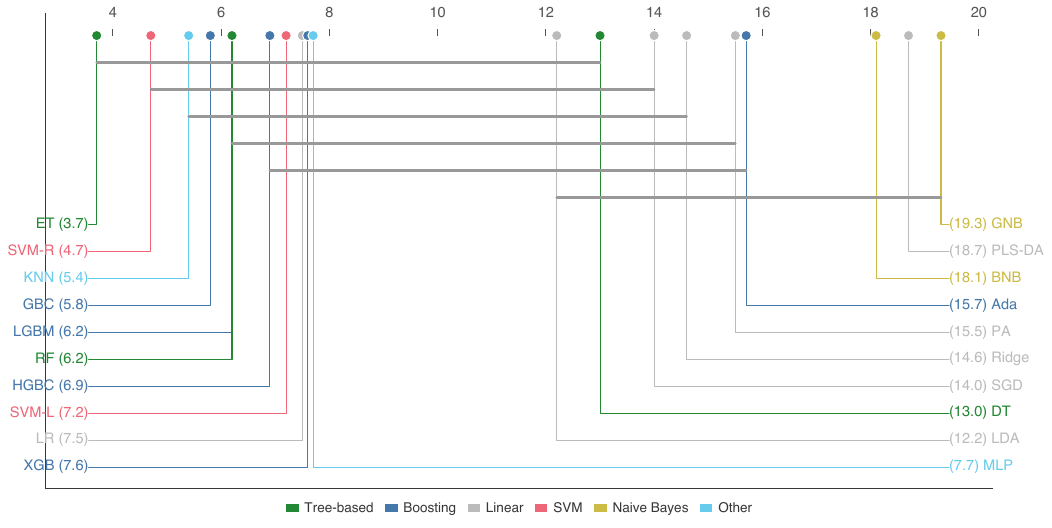}
\caption{Critical difference diagram from Nemenyi post-hoc test ($\alpha=0.05$) on 20-model $\times$ 10-fold overall F1 scores. Models connected by a horizontal bar are not significantly different. Models colored by algorithm family.}
\label{fig:cd-diagram}
\end{figure*}

\begin{figure*}[htbp!]
\centering
\includegraphics[width=\textwidth]{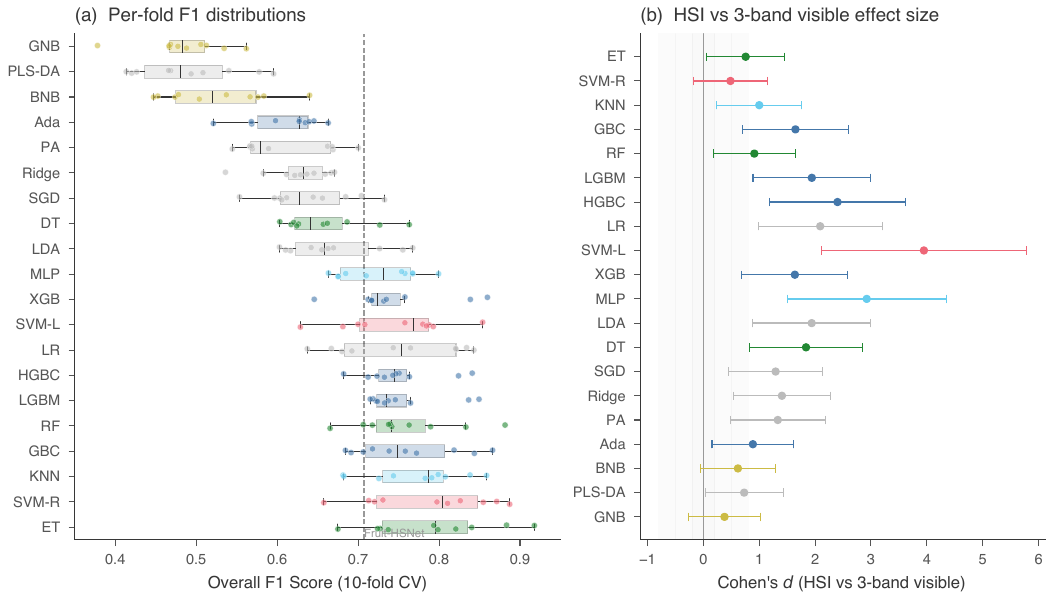}
\caption{(a)~Per-fold overall F1 score distributions from 10-fold stratified cross-validation, sorted by mean F1. Box plots show median and interquartile range; individual fold scores are overlaid as jittered points. Dashed line marks the Fruit-HSNet benchmark (70.7\%). (b)~Per-model Cohen's $d$ effect sizes for the HSI vs.\ VIS-3 subset modality comparison, with 95\% confidence intervals. Shaded bands indicate negligible ($|d|<0.2$), small ($0.2$--$0.5$), and medium ($0.5$--$0.8$) effect regions. Models colored by algorithm family.}
\label{fig:fold-dist}
\end{figure*}

\begin{figure*}[htbp!]
\centering
\includegraphics[width=\textwidth]{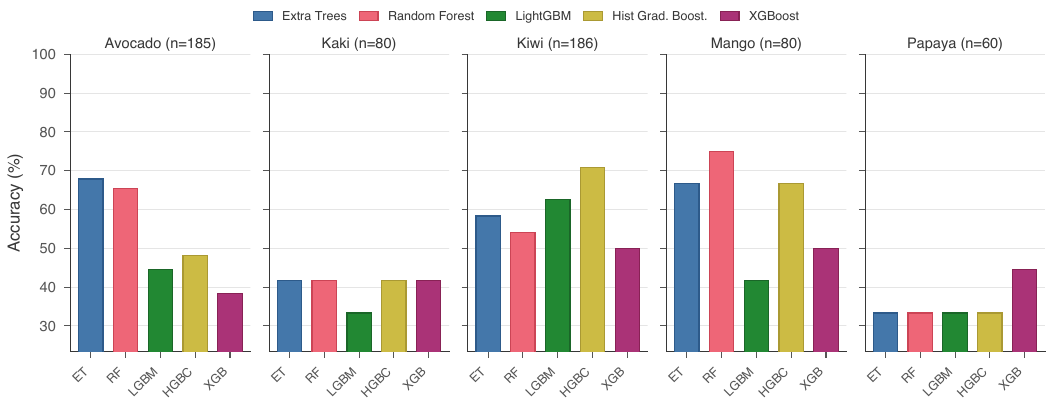}
\caption{Per-fruit overall accuracy for the top-five models (Phase~4). Avocado and mango show highest performance; papaya consistently underperforms due to smallest sample size ($n$=51). Performance variation across species highlights deployment considerations for multi-fruit sorting systems.}
\label{fig:per_fruit}
\end{figure*}

\section{Supplementary Tables}
\label{sec:appendix}

\begin{table*}[htbp!]
\centering
\caption{Overview of 20 evaluated models with Optuna hyperparameter search spaces. HP = number of tuned hyperparameters; FI = native feature importance. All ranges define the Bayesian optimization bounds (100 trials per model).}
\label{tab:model-overview}
\scriptsize
\setlength{\tabcolsep}{3pt}
\resizebox{\textwidth}{!}{%
\begin{tabular}{@{}llcclp{10.5cm}@{}}
\toprule
Abbr. & Model & HP & FI & Lib. & Optuna Search Space \\
\midrule
  ET & Extra Trees & 5 & \checkmark & sklearn & n\_est$\in$[50,500], depth$\in$[3,30], split$\in$[2,20], leaf$\in$[1,20], crit$\in$\{gini,ent\} \\
  RF & Random Forest & 6 & \checkmark & sklearn & n\_est$\in$[50,500], depth$\in$[3,30], split$\in$[2,20], leaf$\in$[1,20], feat$\in$\{sqrt,log2,$\emptyset$\}, crit$\in$\{gini,ent\} \\
  DT & Decision Tree & 4 & \checkmark & sklearn & depth$\in$[3,30], split$\in$[2,20], leaf$\in$[1,20], crit$\in$\{gini,ent\} \\
\midrule
  XGB & XGBoost & 9 & \checkmark & xgboost & n\_est$\in$[50,500], lr$\in$[.01,.3], depth$\in$[3,10], child\_w$\in$[1,10], $\gamma$$\in$[0,1], sub$\in$[.5,1], col$\in$[.5,1], $\alpha$$\in$[0,1], $\lambda$$\in$[0,1] \\
  LGBM & LightGBM & 8 & \checkmark & lightgbm & n\_est$\in$[50,500], lr$\in$[.01,.3], depth$\in$[3,15], leaves$\in$[10,300], child$\in$[5,100], $\alpha$$\in$[0,1], $\lambda$$\in$[0,1], col$\in$[.5,1] \\
  GBC & Gradient Boosting & 5 & \checkmark & sklearn & n\_est$\in$[50,500], lr$\in$[.01,.3], depth$\in$[3,10], split$\in$[2,20], leaf$\in$[1,20] \\
  HGBC & HistGradBoosting & 5 & -- & sklearn & lr$\in$[.01,.3], iter$\in$[100,500], depth$\in$[3,10], leaf$\in$[10,100], L2$\in$[0,1] \\
  Ada & AdaBoost & 3 & \checkmark & sklearn & n\_est$\in$[25,200], lr$\in$[.1,1.5], alg=SAMME \\
\midrule
  LR & Logistic Regression & 3 & -- & sklearn & C$\in$[.01,100], pen$\in$\{L1,L2\}, solv$\in$\{liblinear,saga\} \\
  Ridge & Ridge Classifier & 2 & -- & sklearn & $\alpha$$\in$[.1,100], solver$\in$\{auto,svd,cholesky,lsqr,sag,saga\} \\
  SGD & SGD Classifier & 4 & -- & sklearn & $\alpha$$\in$[1e-6,.01], pen$\in$\{L1,L2\}, lr$\in$\{const,opt,inv,adapt\}, $\eta_0$$\in$[.01,1] \\
  PA & Passive-Aggressive & 3 & -- & sklearn & C$\in$[.01,10], iter$\in$[100,2k], tol$\in$[1e-5,.01] \\
  LDA & Linear Discriminant & 2 & -- & sklearn & solver$\in$\{svd,lsqr,eigen\}, shrink$\in$\{$\emptyset$,auto,.1,.5,.9\} \\
\midrule
  SVM-L & SVC (Linear) & 3 & -- & sklearn & C$\in$[.01,100], tol$\in$[1e-5,.01], iter$\in$[1k,5k] \\
  SVM-R & SVC (RBF) & 3 & -- & sklearn & C$\in$[.01,100], tol$\in$[1e-5,.01], $\gamma$$\in$[1e-6,.1] \\
\midrule
  GNB & Gaussian NB & 1 & -- & sklearn & var\_smooth$\in$[1e-10,1e-6] \\
  BNB & Bernoulli NB & 2 & -- & sklearn & $\alpha$$\in$[.1,10], fit\_prior$\in$\{T,F\} \\
\midrule
  KNN & k-NN & 4 & -- & sklearn & k$\in$[3,50], w$\in$\{uniform,dist\}, alg$\in$\{auto,ball,kd,brute\}, p$\in$\{1,2\} \\
\midrule
  MLP & MLP & 4 & -- & sklearn & hidden$\in$\{(50),(100),(50,50),(100,50),(100,100)\}, $\alpha$$\in$[1e-4,.1], lr$\in$[.001,.1], act$\in$\{relu,tanh,logistic\} \\
\midrule
  PLS-DA & PLS Discriminant Analysis & 1 & -- & sklearn & n\_comp$\in$[1,30] (clamped to $\min(n_\mathrm{samples}, n_\mathrm{features}, n_\mathrm{classes})$) \\
\bottomrule
\end{tabular}%
}
\end{table*}

\begin{table*}[htbp!]
\centering
\caption{Best hyperparameters found by Optuna (100 Bayesian TPE trials per model). Parameters listed after stripping the \texttt{classifier\_\_} prefix. Models ranked by Phase~3 overall accuracy.}
\label{tab:best-hyperparams}
\footnotesize
\setlength{\tabcolsep}{4pt}
\begin{tabular}{clp{11cm}}
\toprule
Rk & Model & Best Hyperparameters \\
\midrule
  1 & ExtraTrees & \texttt{n\_estimators=277, max\_depth=14, min\_samples\_split=6, min\_samples\_leaf=1, criterion=gini} \\
  2 & RandomForest & \texttt{n\_estimators=285, max\_depth=10, min\_samples\_split=5, min\_samples\_leaf=2, max\_features=log2, criterion=entropy} \\
  3 & LGBM & \texttt{n\_estimators=470, learning\_rate=0.2874, max\_depth=14, num\_leaves=10, min\_child\_samples=90, reg\_alpha=0.2373, reg\_lambda=0.4189, colsample\_bytree=0.6112} \\
  4 & HistGradientBoosting & \texttt{learning\_rate=0.1795, max\_iter=123, max\_depth=5, min\_samples\_leaf=41, l2\_regularization=0.5475} \\
  5 & XGBoost & \texttt{n\_estimators=400, learning\_rate=0.1357, max\_depth=10, min\_child\_weight=8, gamma=0.1722, subsample=0.7181, colsample\_bytree=0.5824, reg\_alpha=0.3973, reg\_lambda=0.5746} \\
  6 & KNeighbors & \texttt{n\_neighbors=3, weights=distance, algorithm=ball\_tree, p=1} \\
  7 & GradientBoosting & \texttt{n\_estimators=181, learning\_rate=0.1839, max\_depth=8, min\_samples\_split=11, min\_samples\_leaf=17} \\
  8 & AdaBoost & \texttt{n\_estimators=198, learning\_rate=0.9135, algorithm=SAMME} \\
  9 & GaussianNB & \texttt{var\_smoothing=3.15e-09} \\
  10 & DecisionTree & \texttt{max\_depth=8, min\_samples\_split=17, min\_samples\_leaf=8, criterion=entropy} \\
  11 & SGDClassifier & \texttt{penalty=l2, alpha=5.18e-04, learning\_rate=adaptive, eta0=0.3370} \\
  12 & BernoulliNB & \texttt{alpha=0.1026, fit\_prior=True} \\
  13 & SVC\_RBF & \texttt{C=30.96, tol=1.17e-05, gamma=0.0019} \\
  14 & MLPClassifier & \texttt{hidden\_layer\_sizes=(100, 50), alpha=0.0255, learning\_rate\_init=0.0017, activation=relu} \\
  15 & PassiveAggressive & \texttt{C=3.85, max\_iter=369, tol=0.0092} \\
  16 & SVC\_Linear & \texttt{C=0.0880, tol=1.56e-05, max\_iter=2560} \\
  17 & LogisticRegression & \texttt{penalty=l2, C=35.64, solver=liblinear} \\
  18 & Ridge & \texttt{alpha=93.90, solver=cholesky} \\
  19 & LDA & \texttt{solver=lsqr, shrinkage=0.1000} \\
  20 & PLSDA & \texttt{n\_components=2} \\
\bottomrule
\end{tabular}
\end{table*}

\begin{table}[p]
\centering
\caption{Hardware and Software Environment Specifications. All 20 models were trained on CPU only; no GPU acceleration was used.}
\label{tab:hardware}
\footnotesize
\begin{tabularx}{\textwidth}{@{} l X @{}}
\toprule
\textbf{Component} & \textbf{Specification} \\
\midrule
\addlinespace[0.3em]
\multicolumn{2}{l}{\textit{\textbf{Hardware}}} \\
\addlinespace[0.2em]
\midrule
Processor & Apple M4 (10-core CPU: 4 performance + 6 efficiency) \\
Memory & 24\,GB unified LPDDR5X \\
Storage & SSD \\
GPU & None (CPU-only training) \\
\midrule
\addlinespace[0.3em]
\multicolumn{2}{l}{\textit{\textbf{Software Environment}}} \\
\addlinespace[0.2em]
\midrule
Operating System & macOS (arm64) \\
Python & 3.13 \\
scikit-learn & 1.7.1 \\
XGBoost & 3.0.4 \\
LightGBM & 4.6.0 \\
Optuna & 4.5.0 \\
imbalanced-learn & 0.14.0 \\
SHAP & 0.48.0 \\
LIME & 0.2.0.1 \\
\midrule
\addlinespace[0.3em]
\multicolumn{2}{l}{\textit{\textbf{Reproducibility Settings}}} \\
\addlinespace[0.2em]
\midrule
Global random seed & 42 \\
Seeded components & Python, NumPy, all classifiers, SMOTE, splits, Optuna TPE sampler \\
Data split & Dataset-provided VIS benchmark split (73.4/26.6) \\
Cross-validation & StratifiedKFold, $k{=}10$, shuffle=True \\
Optuna trials & 100 per model (TPE sampler, seed=42) \\
Thread pinning & OMP/OpenBLAS/MKL = 1 thread \\
\bottomrule
\end{tabularx}
\end{table}

\section*{Data Availability}

DeepHS Fruit dataset (version 2) \citep{varga2021measuring} is available at \url{https://github.com/cogsys-tuebingen/deephs_fruit}. Code and supplementary materials will be made available upon publication.

\section*{Declaration of Competing Interest}

The authors declare that they have no known competing financial interests or personal relationships that could have appeared to influence the work reported in this paper.

\section*{CRediT Authorship Contribution Statement}

\textbf{Phongsakon Mark Konrad:} Conceptualization, Methodology, Software, Validation, Formal analysis, Investigation, Data curation, Writing -- original draft, Visualization.
\textbf{Casper Kunstmann-Olsen:} Supervision, Writing -- review \& editing.
\textbf{Jacek Fiutowski:} Supervision, Writing -- review \& editing.
\textbf{Serkan Ayvaz:} Conceptualization, Supervision, Writing -- review \& editing, Project administration.

\section*{Acknowledgements}

This research did not receive any specific grant from funding agencies in the public, commercial, or not-for-profit sectors.

\section*{Declaration of Generative AI and AI-assisted Technologies in the Writing Process}

During the preparation of this work the authors used Claude Opus 4.6 and Claude Sonnet 4.6 (Anthropic) for writing assistance and code development, and Claude Haiku 4.5 (Anthropic) for routine code formatting tasks. After using these tools, the authors reviewed and edited the content as needed and take full responsibility for the content of the published article.

%% Bibliography
\bibliographystyle{elsarticle-harv}
\bibliography{cas-refs}

\end{document}